\documentclass[12pt]{article}
\usepackage{graphicx,cite,epsfig}
\newcommand{\ba}{\begin{array}}
\newcommand{\ea}{\end{array}}
\newcommand{\bd}{\begin{displaymath}}
\newcommand{\ed}{\end{displaymath}}
\newcommand{\be}{\begin{equation}}
\newcommand{\ee}{\end{equation}}
\def\bt{\begin{table}}
\def\et{\end{table}}
\def\bc{\begin{center}}
\def\ec{\end{center}}
\def\bi{\begin{itemize}}
\def\ei{\end{itemize}}
\def\bea{\begin{eqnarray}}
\def\eea{\end{eqnarray}}
\def\beas{\begin{eqnarray*}}
\def\eeas{\end{eqnarray*}}

\def\N0{\widetilde{\chi}^0}

\def\Dm{\widetilde{\Delta}^{--}}

\def\Cm{\widetilde{\chi}^-}
\def\Cpm{\widetilde{\chi}^\pm}

\def\D{\Delta}

\def\slash {\!\!\!\!/}

\catcode`@=11 
\def \gsim{\mathrel{\mathpalette\@versim>}}
\def \lsim{\mathrel{\mathpalette\@versim<}}
\def \@versim#1#2{\lower0.4ex\vbox{\baselineskip\z@skip\lineskip\z@skip
     \lineskiplimit\z@\ialign{$\m@th#1\hfil##\hfil$%
     \crcr#2\crcr\sim\crcr}}} 
\catcode`@=12 
\begin{document} 
\setcounter{page}{0} 
\thispagestyle{empty} 
\begin{flushright} HIP-2007-53/TH \\ CUMQ/HEP 147 \end{flushright} 
\begin{center}
{\Large Single Production of Doubly Charged Higgsinos at linear $e^- e^-$
colliders} 
\\ \vspace*{0.2in} 
{\large Mariana Frank$^{(1)\dagger}$, Katri Huitu$^{(2)\ddagger }$ {\rm
and} Santosh Kumar Rai$^{(2)\ast}$} \\
\vspace*{0.2in}
{\sl $^{(1)}$Department of Physics, Concordia University,\\
7141 Sherbrooke St. West, Montreal, Quebec, Canada H4B 1R6\\ $^{(2)}$ High
     Energy Physics Division, Department of Physical Sciences, University
     of Helsinki,\\ and Helsinki Institute of Physics, P.O. Box 64,
     FIN-00014 University of Helsinki, Finland\\ \rm }

\end{center}
\vspace*{0.6in}
{\large\bf Abstract}

Several extended supersymmetric models, motivated by either grand
unification, or by neutrino mass generation, predict light doubly charged
higgsinos. We study the production of a single doubly charged higgsino and
its decay channels at the International Linear Collider (ILC) operating in
the $e^-e^-$ mode. We analyze the production cross section for $e^-e^- \to
{\tilde \Delta}^{--}_{L,R} \chi^0_1$ as a function of different kinematic
variables, followed by the decay, through several channels, of the doubly 
charged higgsino into a final state of two leptons plus missing energy.
We include the standard model background and discuss how kinematic cuts
could be used effectively to limit this background. Single production of
these exotics could provide a spectacular signal for a new underlying
symmetry and for physics beyond the minimal supersymmetric standard model.


\begin{quotation} 
\noindent \sl
\end{quotation} 
\rm\normalsize
\vfill
$^{\dagger}$mfrank@alcor.concordia.ca

\noindent
$^{\ddagger}$katri.huitu@helsinki.fi

\noindent $^{\ast}$santosh.rai@helsinki.fi 
\newpage 
\section{Introduction}
It is widely expected that the standard model (SM), though successful at
predicting almost all experimental data in low energy physics, is not the
complete theory of fundamental interactions. In addition to the
theoretical inconsistencies and incompleteness of the theory, there have
been recently experimental incentives to study models beyond the SM.  
Observations and measurements of solar and atmospheric neutrino
oscillations \cite{Eidelman:2004wy}, as well as indications of hot and
cold dark matter \cite{Spergel:2003cb} are not predicted, and cannot be
explained, respectively, by the SM.

Supersymmetry, in the form of the minimal supersymmetric standard model
(MSSM) is the most popular scenario of physics beyond the SM. It provides a
satisfactory (at least qualitatively) explanation for dark matter, but
suffers from the same problems as the SM when it comes to explaining
neutrino masses. One must either invoke R-parity violation
\cite{Barbier:2004ez} and abandon the dark matter candidate, or add
right-handed neutrinos and introduce the see-saw mechanism
\cite{Mohapatra:1979ia} . Supersymmetric grand unified theories (SUSY
GUTs)  which contain left-right supersymmetry resolve both problems
naturally \cite{history,Demir:2006ef}. An added bonus of SUSY GUTs is that
electromagnetic, weak and strong interactions unify at the same energy
scale. If such SUSY GUTs containing left-right supersymmetry are present
in nature, one must devise methods to search for signals inherent in them,
and absent in other models. One such signal would be the discovery of
exotic particles unique to such models.

Supersymmetric left-right theories (LRSUSY), based on the product group
$SU(2)_L \times SU(2)_R \times U(1)_{B-L}$, are attractive for many
reasons \cite{history, Francis:1990pi}.  They disallow explicit R-parity
breaking in the Lagrangian; they provide a natural mechanism for
generating neutrino masses; and they provide a solution to the strong and
weak CP problem in MSSM \cite{Mohapatra:1995xd}. Neutrino masses are
induced by the see-saw mechanism through the introduction of Higgs triplet
fields which transform as the adjoint of the $SU(2)_R$ group and have
quantum numbers $B-L=\pm 2$ (where $B$ is baryon, and $L$ is lepton
number).  While the Higgs triplet bosons are present in the
non-supersymmetric version of the theory, their fermionic partners, the
higgsinos, are specific to the supersymmetric version.  It has been shown
that, if the scale for left-right symmetry breaking is chosen so that the
light neutrinos have the experimentally expected masses, these higgsinos
can be light, with masses in the range of ${\cal O} (100)$ GeV
\cite{Chacko:1997cm}. Such particles could be produced in abundance at
future colliders and thus give definite signs of left-right symmetry at
future colliders.

While the production of doubly charged Higgs bosons in the framework of
the left-right model has been investigated at linear accelerators
\cite{Barenboim:1996pt} and at LHC \cite{Huitu:1996su}, the corresponding
higgsinos have received less attention.  The exceptions are the references
in \cite{Huitu:1995bc}, where some of the properties of doubly charged
higgsinos have been highlighted, and in \cite{Raidal:1998vi}, where the
pair production of higgsinos in $e^+e^-$ was analyzed.  We note that
doubly charged higgsinos also appear in the so-called 3-3-1 models (models
based on the $SU(3)_c \times SU(3)_L \times U(1)_{N}$ symmetry)  
\cite{Singer:1980sw}.

In this present work, we concentrate on a definite signal for 
 doubly charged higgsinos:  the production of a single one at
an $e^-e^-$ collider, followed by the decay (through several channels)  to
$e^-e^-$ + $\rm E_{miss}$. In order to obtain definite predictions for the signal, we
perform our analysis in the context of the LRSUSY, though we expect the
results for the 3-3-1 model to be similar. We concentrate first on the
details and characteristics of the production cross section. We then discuss the
possible decay modes of the doubly charged higgsinos (either two-body or
three-body, depending on the spectrum characteristics). We complete our
analysis with the discussion of the SM background, and indicate how cuts
could be employed most efficiently to reduce these backgrounds.

Our paper is organized as follows.  In Section \ref{sec:model} we describe
the LRSUSY model, with particular emphasis on the sectors of interest, as well as summarize the restrictions on the relevant  Yukawa couplings.  
In Section \ref{sec:results} we present the analysis for the production
and decay, separately for the left and right-handed doubly charged higgsino and of the SM backgrounds and cuts needed to observe the signal.  We reach the conclusions in Section \ref{sec:summ}. The Appendix contains the mixing in the  scalar and gaugino/higgsino sectors which enter our calculation.

\section{ Description of the LRSUSY Model}\label{sec:model}
The minimal supersymmetric left-right model is based on the gauge group
$SU(3)_C \times SU(2)_L \times SU(2)_R \times U(1)_{B-L}$. The matter
fields of this model consist of three families of quark and lepton chiral
superfields which transform under the gauge group as:
\begin{eqnarray}
Q&=&\left (\begin{array}{c}
u\\ d \end{array} \right ) \sim \left ( 3,2, 1, \frac13 \right ),~~
Q^c=\left (\begin{array}{c}
d^c\\u^c \end{array} \right ) \sim \left ( 3^{\ast},1, 2, -\frac13
\right ),\nonumber \\
L&=&\left (\begin{array}{c}
\nu\\ e\end{array}\right ) \sim\left ( 1,2, 1, -1 \right ),~~
L^c=\left (\begin{array}{c}
e^c \\ \nu^c \end{array}\right ) \sim \left ( 1,1, 2, 1 \right ),
\end{eqnarray}
where the numbers in the brackets denote the quantum numbers under
$SU(3)_C \times SU(2)_L \times SU(2)_R \times U(1)_{B-L}$. The Higgs
sector consists of the bidoublet and triplet Higgs superfields:
\begin{eqnarray}
&&\Phi_1 = \left (\begin{array}{cc}
\Phi^0_{11}&\Phi^+_{11}\\ \Phi_{12}^-& \Phi_{12}^0
\end{array}\right) \sim \left (1,2,2,0 \right),~~~ 
\Phi_2=\left (\begin{array}{cc}
\Phi^0_{21}&\Phi^+_{21}\\ \Phi_{22}^-& \Phi_{22}^0
\end{array}\right) \sim \left (1,2,2,0 \right) \nonumber \\
&&\Delta_{L}=\left(\begin{array}{cc}
\frac {1}{\sqrt{2}}\Delta_L^-&\Delta_L^0\\
\Delta_{L}^{--}&-\frac{1}{\sqrt{2}}\Delta_L^-
\end{array}\right) \sim (1,3,1,-2),~~
\delta_{L}  =
\left(\begin{array}{cc}
\frac {1}{\sqrt{2}}\delta_L^+&\delta_L^{++}\\
\delta_{L}^{0}&-\frac{1}{\sqrt{2}}\delta_L^+
\end{array}\right) \sim (1,3,1,2),\nonumber \\
&&\Delta_{R} =
\left(\begin{array}{cc}
\frac {1}{\sqrt{2}}\Delta_R^-&\Delta_R^0\\
\Delta_{R}^{--}&-\frac{1}{\sqrt{2}}\Delta_R^-
\end{array}\right) \sim (1,1,3,-2),~~
\delta_{R}  =
\left(\begin{array}{cc}
\frac {1}{\sqrt{2}}\delta_R^+&\delta_R^{++}\\
\delta_{R}^{0}&-\frac{1}{\sqrt{2}}\delta_R^+
\end{array}\right) \sim (1,1,3,2)
\end{eqnarray}
The bi-doublet Higgs superfields $\Phi_1, \Phi_2$ break
the $SU(2)_L \times U(1)_{Y}$ symmetry and generate
a Cabibbo-Kobayashi-Maskawa mixing matrix. Supplementary Higgs
representations are needed to break left-right symmetry spontaneously:
 triplet Higgs
$\Delta_L, ~\Delta_R$ bosons are chosen to also support the seesaw
mechanism. Since the theory is supersymmetric, additional
triplet superfields $\delta_L, ~\delta_R$ are needed to cancel triangle
gauge anomalies in the fermionic sector. The most general superpotential
involving these superfields is:
\begin{eqnarray}
\label{superpotential}
W & = & {\bf Y}_{Q}^{(i)} Q^T\Phi_{i}i \tau_{2}Q^{c} + {\bf Y}_{L}^{(i)}
L^T \Phi_{i}i \tau_{2}L^{c} + i({\bf h}_{ll}L^T\tau_{2} \delta_L L +
{\bf h}_{ll}L^{cT}\tau_{2}
\Delta_R L^{c}) \nonumber \\
& & + \mu_{3}\left [Tr (\Delta_L  \delta_L +\Delta_R 
\delta_R)\right] + \mu_{ij}Tr(i\tau_{2}\Phi^{T}_{i} i\tau_{2} \Phi_{j})
+W_{NR} 
\end{eqnarray}
where $W_{NR}$ denotes (possible) non-renormalizable terms arising from
higher scale physics or Planck scale effects \cite{Chacko:1997cm}. The
presence of these terms insures that, when the SUSY breaking scale is
above $M_{W_{R}}$, the ground state is R-parity conserving. In addition,
the potential also includes $F$-terms, $D$-terms as well as soft
supersymmetry breaking terms:
\begin{eqnarray}
\label{eq:soft}
&&{\cal L}_{soft}=\left[ {\bf A}_{Q}^{i}{\bf Y}_{Q}^{(i)}{\tilde Q}^T\Phi_{i}
i\tau_{2}{\tilde Q}^{c}+ {\bf A}_{L}^{i}{\bf Y}_{L}^{(i)}{\tilde L}^T \Phi_{i}
i\tau_{2}{\tilde L}^{c} + i{\bf A}_{LR} {\bf h}_{ll}({\tilde L}^T\tau_{2}
\delta_L{\tilde  L} + {\tilde L}^{cT}\tau_{2} \Delta_R{\tilde L}^{c}) 
\right.
\nonumber
\\ & & \left. +{ m}_{\Phi}^{(ij) 2}
\Phi_i^{\dagger}  \Phi_j \right] + \left[( m_{L}^2)_{ij}{\tilde l}_{Li}^{\dagger}{\tilde
l}_{Lj}+ (m_{R}^2)_{ij}{\tilde l}_{Ri}^{\dagger}{\tilde l}_{Rj} \right]- M_{LR}^2 \left[
Tr(  \Delta_R  \delta_R)+ Tr(  \Delta_L 
 \delta_L) +h.c.\right] \nonumber \\
& &- [B \mu_{ij} \Phi_{i} \Phi_{j}+h.c.] 
\end{eqnarray}

The symmetry is broken spontaneously to $U(1)_{em}$ when the neutral Higgs
fields acquire non-zero vacuum expectation values $(VEV's)$:
\begin{eqnarray}
\langle \Phi_1 \rangle = \left (\begin{array}{cc}
\kappa_1&0\\0&\kappa_1^{\prime} e^{i\omega_1}
\end{array}\right),~\langle \Phi_2 \rangle = \left (\begin{array}{cc}
\kappa_2^{\prime}e^{i \omega_2}&0\\0&\kappa_2
\end{array}\right), ~\langle \Delta_{L} \rangle = \left(\begin{array}{cc}
0&v_{\Delta_L}\\0&0
\end{array}\right), \nonumber \\
\langle ~\delta_{L} \rangle = \left
(\begin{array}{cc} 0&0\\v_{\delta_L}&0
\end{array}\right)~,~\langle ~\Delta_{R} \rangle = \left
(\begin{array}{cc} 0&v_{\Delta_R}\\0&0
\end{array}\right), ~\langle ~\delta_{R} \rangle = \left
(\begin{array}{cc} 0&0\\v_{\delta_R}&0
\end{array}\right).
\nonumber
\end{eqnarray}
The Lagrangians in (\ref{superpotential}) and (\ref{eq:soft}) give rise to the interactions of the doubly charged higgsinos $\tilde \Delta_{L,R}^{--}$. As these are determined by the magnitude of the triplet Yukawa couplings, we review first the restrictions on these.

\subsection{Experimental limits on triplet masses and Yukawa couplings}
\label{subsec:constraints}

Indirect experimental limits for the triplet Yukawa couplings come from
lepton number violating processes mediated by doubly charged Higgs bosons.
These processes constrain the first two generations, while the third
generation couplings remain constrained only by the requirement of
perturbativity. The constraints on Yukawa couplings as a function of the
doubly charged Higgs mass can be found in 
\cite{Swartz:1989qz,Willmann:1998gd} and are
given as follows
\bea 
&&h_{e\mu}h_{ee}<3.2\times 10^{-11} \,{\rm
GeV}^{-2}\cdot M_{\Delta^{--}}^2 \quad {\rm from}\,\mu\rightarrow \bar e
ee,\nonumber\\ 
&&h_{e\mu}h_{\mu\mu}<2\times 10^{-10} \,{\rm GeV}^{-2}\cdot
M_{\Delta^{--}}^2 \quad {\rm from}\,\mu\rightarrow e \gamma,\nonumber\\
&&h_{ee}^2<9.7\times 10^{-6} \,{\rm GeV}^{-2}\cdot M_{\Delta^{--}}^2 \quad
{\rm from \, Bhabha \, scattering},\nonumber\\ 
&&h_{\mu\mu}^2<2.5\times
10^{-5} \,{\rm GeV}^{-2}\cdot M_{\Delta^{--}}^2 \quad {\rm from} \quad
(g-2)_\mu,\nonumber\\ 
&&h_{e e}h_{\mu\mu}<2.0\times 10^{-7} \,{\rm
GeV}^{-2}\cdot M_{\Delta^{--}}^2 \quad {\rm from \, muonium-antimuonium \,
transition}. \nonumber 
\eea

Bhabha scattering indirect limits were studied also at LEP, and it was
found that if $h_{ee}>0.7$, the doubly charged Higgs mass should be in the 
TeV region \cite{Abbiendi:2003pr,Achard:2003mv}.
Also at LEP,  direct searches of doubly charged Higgs were performed.
From the pair production of doubly charged Higgses a lower bound around
100 GeV was established, if $h_{ij}>10^{-7}$, $i,j=e,\mu ,\tau$
\cite{Abdallah:2002qj,Achard:2003mv}.
From the single production of doubly charged Higgs
boson in $e^+e^-$ collisions, couplings  
$h_{ee}<0.071$ are allowed, if $M_{\Delta^{--}}< 160$ GeV, assuming 100\%
branching fraction to leptons
\cite{Abbiendi:2003pr}. 
At HERA the single production of doubly
charged Higgs boson was studied in $e^+p$ collisions \cite{Aktas:2006nu}
and a lower limit of 141 GeV was found on the doubly charged Higgs mass 
if the coupling $h_{e\mu}=0.3$, 
while the mass limit for $h_{e\tau}=0.3$ was 112 GeV. For
heavier doubly charged Higgses, the corresponding Yukawa couplings are
less restricted, and for $M_{\Delta^{--}} > 150$ GeV triplet Yukawa
couplings are not constrained by HERA. Note that the direct search limits
are the
only ones which include the third generation couplings. 

For simplicity we
assume that the flavor-violating off-diagonal entries in the coupling
matrix are zero and only the flavor diagonal elements have finite
entries.  
Assuming that the triplet Yukawa couplings are degenerate
the most stringent constraints for the $\D L=2$ couplings come
from results for the muonium-antimuonium transitions which depend on the
mass scale for the triplet Higgs fields {\it viz.} 
$h_{ll} \equiv \tilde{f}_{ll}
< 0.44 M_{\D^{\pm\pm}}~{\rm TeV}^{-1}$, at 90\% C.L. \cite{Willmann:1998gd}.

We now proceed to examine the doubly charged higgsino sector of the theory. 
We first review the known information on their masses and couplings, then 
proceed to discuss their production and decay channels. For completeness, we 
present the information on the mixing in the chargino, neutralino and charged 
scalar sectors of our theory, insofar as relevant to our discussion of doubly
charged higgsinos, in the Appendix.

\section{Production and Decay of Doubly Charged Higgsinos}\label{sec:results}

We analyze first the single production of the doubly charged higgsino at the
next generation linear collider running in the $e^- e^-$ mode. The linear
collider operating in this mode will provide an ideal environment for
single production of such doubly
\begin{figure}[htb]
\begin{center}
\includegraphics[height=2.0in,width=2.5in]{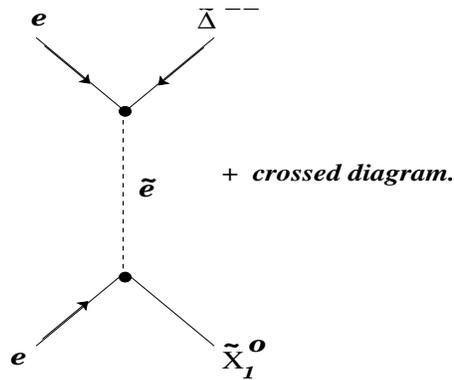}
\caption{\sl\small Feynman graphs for the two-body production of $\Dm \N0_1$
at $e^-e^-$ collider.}
\label{feyngraph}
\end{center}
\end{figure}
charged particles and signals for such processes can be remarkably free 
from backgrounds. We focus on a typical production mode for the doubly charged 
higgsinos $\Dm$ in association with the lightest neutralino and calculate 
the rates for the process: 
\be
e^{-} e^{-} \longrightarrow \Dm \N0_1 
\label{prodeqn}
\ee
and study the signal resulting from the decays of the $\Dm$.  This
production is mediated by the selectron exchange\footnote{We neglect the 
contribution coming from the exchange of
doubly charged Higgs ($\D$) in the s-channel which we find to be negligibly 
small in comparison. This is because of the huge suppression coming from the 
mass of $\D$ in the propagator, which we have assumed to be large.} 
as shown in Fig~\ref{feyngraph} and would be accompanied by large missing 
energy as the lightest neutralino in the final state  is stable and escapes
undetected. The $e^-e^-$ production mode would allow us to probe a large 
range of masses of the doubly charged higgsino, compared to the case where 
one produces them in pairs at linear colliders operating in the $e^+e^-$ mode.
\bt[htb]
$$
\begin{array}{|ll|ll|}\hline
\underline{{\rm Sample~point~{\bf A}}}:&\tan\beta=20 
&M_{\N0_1}=91.8~GeV&M_{\N0_2}=180.6~GeV \\
M_{B-L}= 25~GeV&M_L=M_R=250~GeV
&M_{\N0_3}=250.0~GeV&  \\
\mu_1 =1000~GeV&\mu_3 = 300~GeV& &  \\
v_{\D_L}=1.5\times10^{-8}~GeV&v_{\delta_L}=1.0\times10^{-8}~GeV
&M_{\Cpm_1}=249.0~GeV&M_{\Cpm_2}=300.0~GeV \\
v_{\D_R}=3000~GeV&v_{\delta_R}=1000~GeV
&M_{\Cpm_3}=911.7~GeV& \\\hline\hline
\underline{{\rm Sample~point~{\bf B}}}:&\tan\beta=30 
&M_{\N0_1}=217.3~GeV&M_{\N0_2}=441.7~GeV \\
M_{B-L}=100~GeV&M_L=M_R= 500~GeV
&M_{\N0_3}=450.0~GeV&  \\
\mu_1 = 500~GeV&\mu_3 = 500~GeV& &  \\
v_{\D_L}=1.5\times10^{-8}~GeV&v_{\delta_L}=1.0\times10^{-8}~GeV
&M_{\Cpm_1}=447.8~GeV&M_{\Cpm_2}=500.0~GeV \\
v_{\D_R}=2500~GeV&v_{\delta_R}=1500~GeV               
&M_{\Cpm_3}=500~GeV& \\\hline
\end{array}
$$
\caption{\sl\small Sample points of the particle spectrum. The left column 
contains the input values for the parameters of the model, while the right
column lists the masses of the lightest the neutralinos and charginos corresponding
to the choice of input parameters.}
\label{susyin}
\et
The upper value of the mass of the produced $\Dm$ will be constrained only
by the mass of the lightest neutralino and the kinematic threshold of the
machine energy of the collider. In the latter part of this section we will 
focus on the possible signatures for different final states.

\begin{figure}[htb]
\begin{center}
\includegraphics[height=3.1in,width=3.1in]{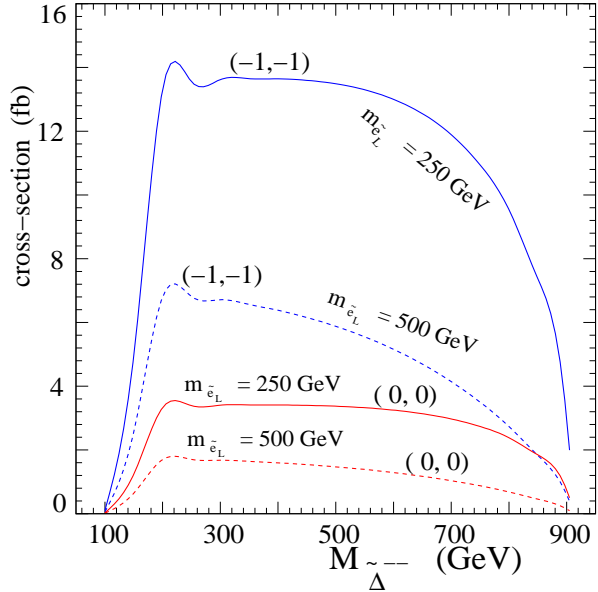}
\caption{\sl\small 
The production cross section as a function of the triplet higgsino
mass for $\sqrt{s}=1$ TeV. We highlight the dependence of the production cross section 
(for $\Dm_L$) on the beam polarization (-1,-1) as compared with the 
unpolarized cross sections for two choices of selectron mass 
(250 GeV and 500 GeV). The plot is produced for the {\it sample point} {\bf A} 
given in Table~\ref{susyin}.}
\label{prodfig}
\end{center}
\end{figure}
For the analysis, we choose two representative points in the parameter 
space of the model as presented in Table~\ref{susyin} 
(we have consistently chosen negligibly small values for the VEV's
$v_{\D_L}$ and $v_{\delta_L}$ so that the constraints on the $\rho$-parameter 
is not disturbed). This will lead, 
as our discussions will show, to clear signals for doubly charged higgsino 
production. Moreover, a reasonable choice for the $\D L=2$ couplings
leads to like-sign lepton signals which can have very suppressed or no
SM background. For simplicity, we have however focused only on the 
diagonal $\D L=2$ couplings. 
As one could produce both the left or right-chiral states of $\Dm$, we
divide this section into two parts and discuss the production of
the two chiral states separately.

\subsection{The left-handed higgsino $\Dm_L$}
\subsubsection{Production}\label{subsec:prodLL}
In Fig.~\ref{prodfig} we plot the production cross sections of 
$\Dm_L$ with both 
unpolarized and polarized $e^-$ beams.  
The center-of-mass energy is taken to be
$\sqrt{s}=1$ TeV\footnote{While the initial center-of-mass energy at the
ILC \cite{Heuer:2006ka} is expected to be $\sqrt{s}=500$ GeV, it is likely 
to be increased soon to 1 TeV, for which more promising signals can be 
observed.}. We choose a coupling
strength of $\tilde{f}_{ee}=0.1$,  in agreement with bounds on
$\D L=2$ couplings from different experimental data as listed in section
~\ref{subsec:constraints}.
With masses for the triplet Higgs fields in the TeV-range one can easily
obtain larger values for the couplings. In contrast to existing bounds on
the mass
of the triplet Higgs (quite severe for larger values of the $\D L=2$
couplings), the mass of $\Dm_L$ is not strongly constrained
($\gsim$ 100 GeV). So a light $\Dm_L$ with a large $\D L=2$
coupling
is not ruled out and remains consistent with experimental data. In
the sample points given in Table~\ref{susyin} the lightest neutralino is
the
lightest supersymmetric particle (LSP). We have used
{\it sample point} {\bf A} in calculating the rates in Fig.~\ref{prodfig}.

The production cross sections are shown for two different choices of selectron 
mass {\it viz.} 250 GeV and 500 GeV. As expected, the larger value of the 
selectron mass in the propagator suppresses the production cross section.  
The production is however enhanced by a factor of four for the left-chiral 
$\Dm_L$, if one uses 100\% left-polarization for both beams, as shown in 
Fig~\ref{prodfig}. Other combinations make the production cross section 
vanish for the left-chiral state because of the $(1-\gamma_5)$ coupling (see
Section~\ref{subsec:couplings}). 
The figure shows that even with a conservative value for the $\D L=2$
coupling, one expects a production cross section between 10-14 fb for polarized
beams and a 250 GeV-mass selectron. With an integrated luminosity of 
500 $fb^{-1}$ one can have a large production rate of the doubly charged 
higgsino $\Dm_L$.

\begin{figure}[htb]
\begin{center}
\includegraphics[height=2.8in,width=2.8in]{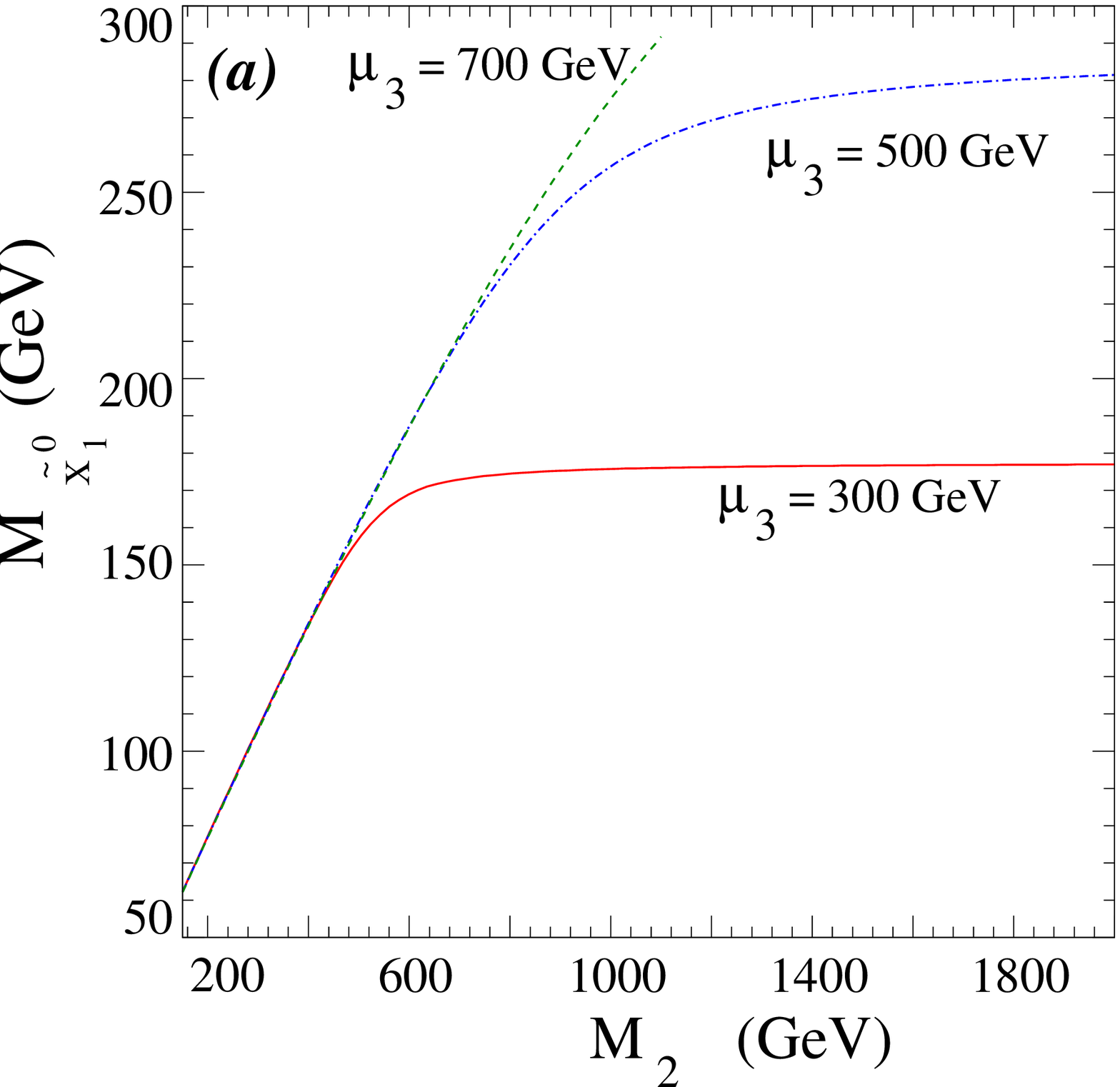}
\includegraphics[height=2.8in,width=2.8in]{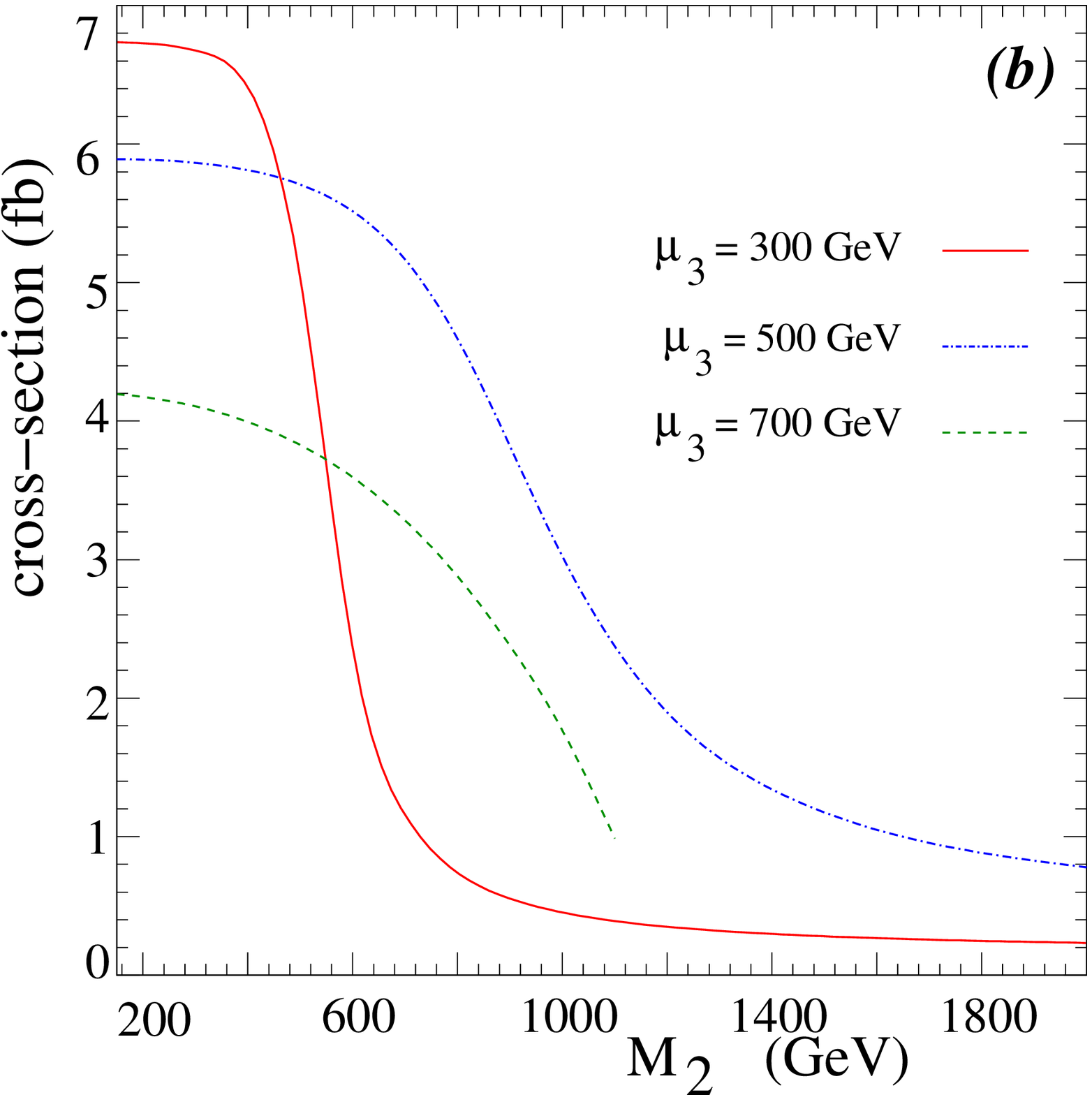}
\caption{\sl\small (a) The lightest neutralino mass as a function of 
$M_L=M_R=M_2$. (b) The production cross section as a 
function of $M_2$, for three different values of $\mu_3$. The other parameters 
are matched with the remaining inputs to {\it sample point} {\bf A} given in 
Table~\ref{susyin}.
Here $\sqrt{s}=1$ TeV, $m_{\tilde e_L}=$ 500 GeV and 
$M_{\tilde\Delta_L^{--}}=\mu_3$.}
\label{ML}
\end{center}
\end{figure}

It is worth noting that the choice of the soft parameters have a role to play 
in the production mechanism, as they form the basic entries in the matrix 
diagonalizing the neutralino mass matrix (see Section~\ref{subsec:neutralinos}). 
The $e-\tilde{e}-\N0$ coupling, which can be written down as 
\beas
&&\ell^-{\tilde \ell}_L^{-} \chi_k^0~~\to~~\frac{1}{\sqrt{2}}(g_L N_{k1} 
+ g_V N_{k3}) P_L
\eeas  
and the components of the lightest neutralino mass eigenstate will
thus be modified for different choices of the soft parameters in the theory. 
A quick look at Table~\ref{susyin} shows that the mass splitting between the
lightest neutralino and the next lightest neutralino strongly depends on the
ratio $\frac{v_{\D_R}}{v_{\delta_R}}$ where a larger ratio results in smaller
mass splitting. This can lead to interesting signatures where the next lightest
neutralino can become the next-lightest SUSY particle (NLSP). Such interesting 
\begin{figure}[htb]
\begin{center}
\includegraphics[height=2.8in,width=2.8in]{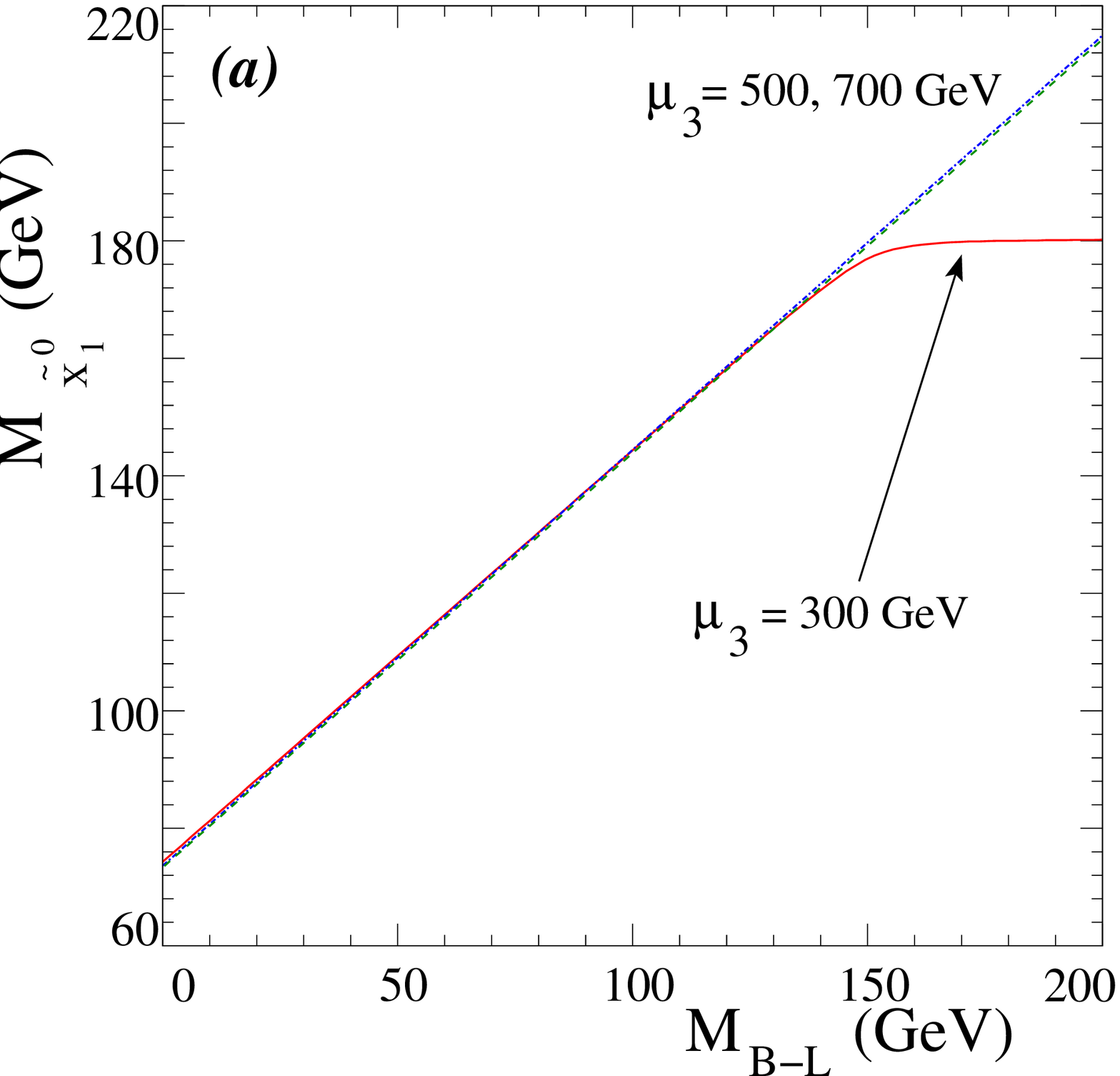}
\includegraphics[height=2.8in,width=2.8in]{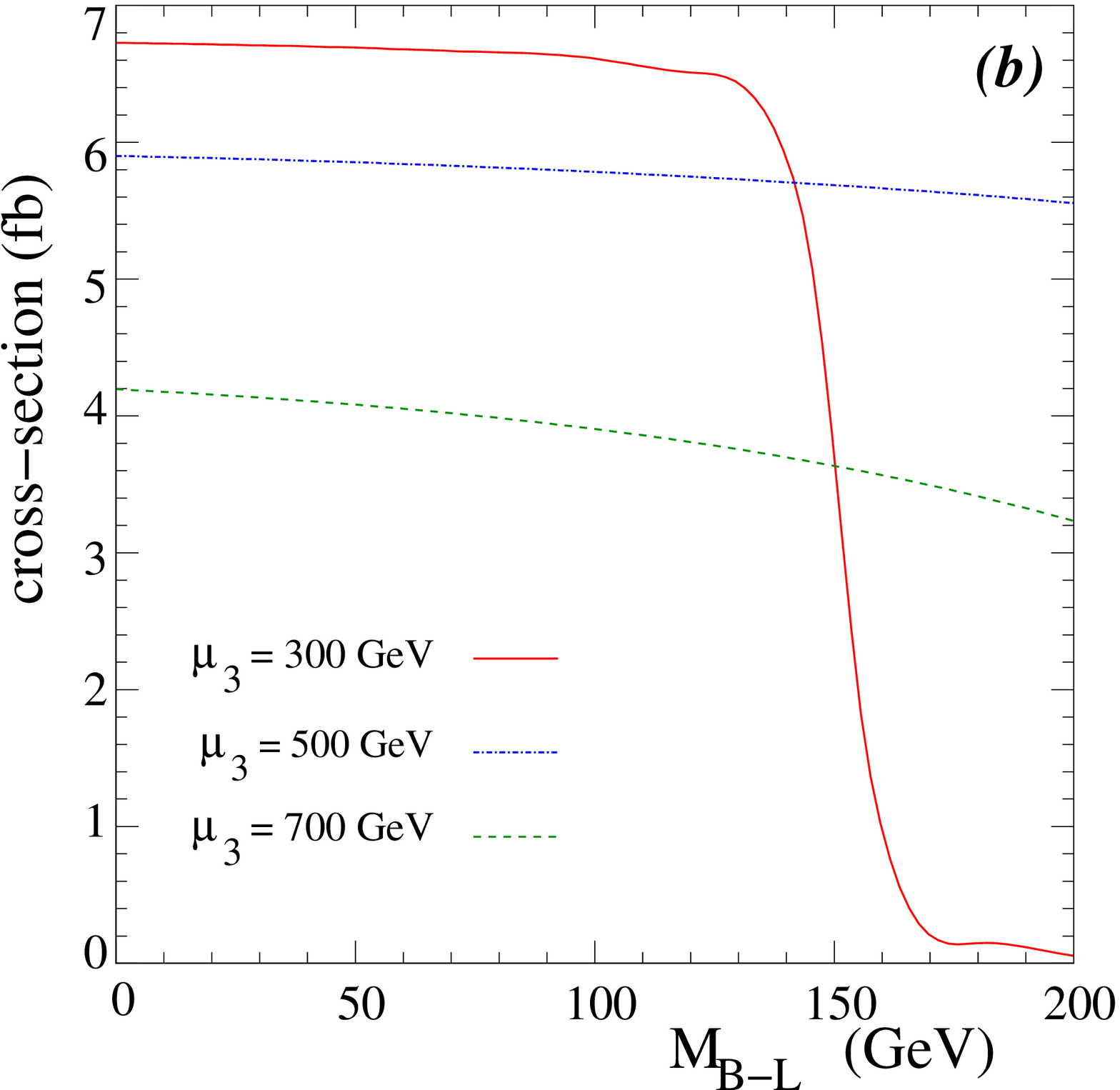}
\caption{\sl\small (a) The lightest neutralino mass as a function of 
$M_{B-L}$. (b) The  production cross section as a function of 
$M_{B-L}$, for three different values of $\mu_3$. 
The other parameters are matched with the remaining inputs 
to {\it sample point} {\bf A} given in Table~\ref{susyin}. 
Here $\sqrt{s}=1$ TeV, $m_{\tilde e_L}=$ 500 GeV and 
$M_{\tilde\Delta_L^{--}}=\mu_3$.}
\label{MV}
\end{center}
\end{figure}
spectrum analysis is however left for future work and not considered here 
anymore. We show the dependence of the relevant soft parameters on the lightest 
neutralino mass and the corresponding production cross sections in 
Fig.~\ref{MV} and Fig.~\ref{ML}. The mass of the doubly-charged 
higgsino is $M_{\Dm_L}=\mu_3$, with the exchanged selectron mass 
fixed at $m_{\tilde{e}_L}=500$ GeV while the polarization choice for the 
colliding electron beams is (-1,-1) which maximizes the production 
cross section for the left-handed higgsino. In Fig.~\ref{MV}(a), 
we show the lightest neutralino mass as a function of the $U(1)_{B-L}$ 
gaugino mass parameter $M_{B-L}$ for three different values of $\mu_3$. 
We find that, increasing value of $M_{B-L}$ (the other model parameters fixed to 
values for {\it sample point} {\bf A} given in 
Table~\ref{susyin}), the lightest neutralino state becomes heavier. In fact
the the lightest neutralino mass shows no dependence on $\mu_3$ for
low values of $M_{B-L}$, but can be seen to admit a dominant admixture of
the right higgsino state beyond $M_{B-L}\sim$140 GeV for the lower choice
of $\mu_3=$ 300 GeV. This fact is highlighted in the plot for the 
cross section, which starts falling rapidly for values of $M_{B-L}\ge$ 140 GeV 
as shown in Fig.~\ref{MV}(b), for $\mu_3=$ 300 GeV. 
For large $M_{B-L}$, the lightest state contains 
a dominant admixture of the right higgsino states and thus its coupling strength is 
decreased. The cross section for the other choices of $\mu_3=$ 500 GeV and 700 GeV
do not show the sharp fall as the lightest neutralino remains dominantly 
$U(1)_{B-L}-$gaugino-like for the choice of $M_{B-L}$ in the plots.   

In Fig.~\ref{ML}(a) we plot the lightest neutralino mass as a function of
$M_L=M_R=M_2$ for the three different values of $\mu_3$. The mass of $\N0_1$
again shows little dependence on $\mu_3$ for low values of $M_2$. The 
contribution of the right higgsino state in the lightest neutralino shows
up early for the lower values of $\mu_3$ as seen in Fig.~\ref{ML}(a).
As shown in Fig.~\ref{ML}(b), in contrast to Fig.~\ref{MV}(b), 
the cross section varies over a larger value when increasing the value of the 
neutralino mass (increasing value of the soft parameter). The 
cross section is again seen to show a sharp fall at different values of $M_2$
for different choices of $\mu_3$. This is again due to the increase of the 
right-higgsino component in the lightest neutralino
and the corresponding change in the values of the $e-\tilde{e}_L-\N0_1$ coupling.
The curves for $\mu_3=$ 700 GeV end abruptly, because the cross section
becomes zero as the lightest neutralino mass becomes greater than 300 GeV
and the process is kinematically disallowed at the $\sqrt{s}=$ 1 TeV 
machine.

We  now proceed to perform the signal analysis for the $\Dm_L$ production in
detail.

\subsubsection{Decays}\label{subsec:decayLL}

We focus on the event rates for specific final states arising
through the decay of the left-chiral doubly charged higgsino. This would
require the knowledge of all possible decay modes accessible to the $\Dm$.
In general,  the following 2-body decays of the doubly charged
higgsinos are allowed: 
\begin{itemize} 
\item $\Dm \longrightarrow \widetilde{\ell}^- ~\ell^-$, 
\item $\Dm \longrightarrow \Delta^{--} ~ \N0_i$, 
\item $\Dm \longrightarrow \Cm_i ~ \Delta^{-}$. 
\end{itemize} 

However, we assume the triplet Higgses to be heavier (which allows a less
constrained $\D L=2$ coupling), and degenerate in mass, which renders them
kinematically inaccessible for decay modes of the relatively lighter doubly
charged higgsinos. Hence, we concentrate on the only favored channel for
decay which is $\Dm \longrightarrow \widetilde{\ell}^- ~ \ell^-$, provided
$m_{\tilde{l}} < M_{\Dm}$. For relatively light higgsinos, one can in
principle have $m_{\tilde{l}} > M_{\Dm}$ and in such a scenario the
only allowed decay mode would be the 3-body decays for the doubly charged
higgsinos, which would dominantly be through off-shell sleptons:  $\Dm \to
\widetilde{\ell}^{*-} ~ \ell^- \to \ell^- \ell^- \N0_1$. We focus our analysis
on the final states:  two like signed leptons in association with large
missing energy coming from the LSP ($\N0_1$).

\subsubsection{Analysis of the final states}\label{subsec:analysis}
 
In the rest of the analysis of final states, we will consider the
2-body and 3-body decay modes of the doubly charged higgsino ($\Dm_L$),
and look at the resulting signal events against the most dominant SM
background.  We present our results for both the sample points given in
Table~\ref{susyin} with point {\bf A} used for a machine with
$\sqrt{s}=500$ GeV and point {\bf B}  considered for
$\sqrt{s}=1$ TeV option. In the case of {\it sample point} {\bf A}, the
mass of the doubly charged higgsino ($\Dm_L$) is taken to be 300 GeV and
the LSP mass is 91.8 GeV, while the other neutralino states are
heavy and not accessible to the decay of the doubly charged higgsino
or sleptons (directly). This also makes both the 2-body $\Dm \longrightarrow
\widetilde{\ell}^- ~ \ell^-$ and 3-body decay of $\Dm \to \ell^- \ell^-
\N0_1$  the only allowed channels of decay.  However, we  point out
that if the other decay modes ($\N0_i \ell^-,i>2$ and $\Cpm_i
\nu_{\ell}$) were accessible as the decay channels of sleptons, then the only
visible effect on our signal would be to reduce the branching fractions
and hence decrease the signal events for our final states.  But they will
not change the characteristic kinematic features of the final states in
consideration. In our analysis we focus on the following leptonic final states:

\bc
(i)~~~$e^-e^- E\slash$, ~~~~ (ii)~~~$\mu^-\mu^- E\slash$  
\ec

It is worth pointing out that the analysis for a final state 
$\tau^-\tau^- E\slash$ will be exactly similar to $\mu^-\mu^- E\slash$, 
provided that the $\D L$ couplings are identical and that we assume similar 
detection efficiencies for the $\tau$'s in the final state. Thus we do not 
discuss the final states with $\tau$'s in our analysis any further. 
Assuming, as before, beam polarization of (-1,-1) throughout the analysis for 
the $\Dm_L$ production, the major SM background that contributes to the final 
states in (i) is the scattering process: 
$$e^- + e^- \to  e^- + e^- + \bar{\nu}_l \nu_l$$ 
\bt[hb]
\small{
\bc
\begin{tabular}{|ll|c|c|c|}
\hline
&&\multicolumn{3}{|c|}{$\sqrt{s}=500~\rm{GeV}$}\\\hline
Cuts Used&& SM &signal-1&signal-2\\\hline
$E_e>5$ GeV& $|\eta_{e}|<3.0$
                       &(-,-) 537.7 fb&(-,-) 123.8 fb&(-,-) 19.9 fb\\
$E\slash > 10$ GeV& $\D R_{ee}>0.2$
                       &(+,+) 15.1 fb&(+,+)  486.8 fb&(+,+) 78.1 fb  \\\hline
$E_e>5$ GeV& $|\eta_{e}|<1.5$ 
                       &(-,-) 218.0 fb&(-,-) 102.7 fb&(-,-) 16.5 fb\\
$E\slash > 100$ GeV& $\D R_{ee}>0.2$ 
                       &(+,+) 3.8 fb&(+,+) 403.98 fb&(+,+)  65.1 fb \\\hline
$E_e>5$ GeV& $|\eta_{e}|<3.0$ 
                       &(-,-) 280.1 fb&(-,-) 123.8 fb&(-,-) 19.9 fb\\
$E\slash > \sqrt{s}/2$ GeV& $\D R_{ee}>0.2$
                       &(+,+) 3.3 fb&(+,+)  468.8 fb&(+,+)  78.1 fb \\\hline
$E_e>5$ GeV& $|\eta_{e}|<1.5$
                       &(-,-) 103.8 fb&(-,-) 102.7 fb&(-,-) 16.5 fb\\
$E\slash > \sqrt{s}/2$ GeV& $\D R_{ee}>0.2$
                       &(+,+) 0.103 fb&(+,+) 403.98 fb&(+,+)  65.1 fb \\\hline
&&\multicolumn{3}{|c|}{$\sqrt{s}=1~\rm{TeV}$}\\\hline
$E_e>5$ GeV& $|\eta_{e}|<3.0$
                       &(-,-) 1.13  pb&(-,-) 40.5 fb&(-,-) 14.0 fb \\
$E\slash > 10$ GeV& $\D R_{ee}>0.2$
                       &(+,+) 12.6 fb&(+,+) 156.4 fb&(+,+)  53.9 fb \\\hline
$E_e>5$ GeV& $|\eta_{e}|<1.5$ 
                       &(-,-) 238.9 fb&(-,-) 33.4 fb&(-,-) 11.7 fb \\
$E\slash > 100$ GeV& $\D R_{ee}>0.2$ 
                       &(+,+) 3.1 fb&(+,+) 129.2 fb&(+,+)  45.0 fb \\\hline
$E_e>5$ GeV& $|\eta_{e}|<3.0$ 
                       &(-,-) 605.9 fb&(-,-) 40.5 fb&(-,-) 14.0 fb \\
$E\slash > \sqrt{s}/2$ GeV& $\D R_{ee}>0.2$
                       &(+,+) 0.4 fb&(+,+) 156.4 fb&(+,+)  53.9 fb \\\hline
$E_e>5$ GeV& $|\eta_{e}|<1.5$
                       &(-,-) 106.0 fb&(-,-) 33.4 fb&(-,-) 11.7 fb \\
$E\slash > \sqrt{s}/2$ GeV& $\D R_{ee}>0.2$
                       &(+,+) 0.007 fb&(+,+) 129.2 fb&(+,+)  45.0 fb \\\hline
\end{tabular}
\caption{\sl\small Signal and SM cross sections for the $e^- e^- E\slash$ 
final states with different choice of kinematic cuts for both signal and 
background at the $e^- e^-$ collider with center-of-mass energies 
$\sqrt{s}=500$ GeV and $\sqrt{s}=1$ TeV. 
We also show the beam polarizations in parentheses. 
The $\D R$ is defined as  $(\D R)^2\equiv (\D\phi)^2+(\D\eta)^2$ with $\D\eta$
and $\D\phi$ respectively denoting the separation in rapidity and azimuthal
angle for the pair of particles under consideration.}
\label{rates}
\ec
}
\et
which, although a continuum background, could {\it prima facie} be
very large. Thus, the event selection criteria are largely aimed at 
suppressing this continuum background. The SM background for signal (ii)
comes from a six-body final state 
$$e^- + e^- \to  \mu^- + \mu^- +\bar{\nu}_\mu\bar{\nu}_\mu+\nu_e\nu_e$$ 
and is quite small.

To highlight
the effect of kinematic cuts in suppressing the background, we present some 
results in Table~\ref{rates}. The SM background has been calculated 
using the event generation package of Madgraph and Madevent 
\cite{Maltoni:2002qb},
with slight modifications to extract the polarized cross sections. 
We have chosen the $|\D L|=2$ coupling strength as $\tilde{f}_{ee}=0.3$ 
throughout the analysis and the branching ratios for the 2-body and 3-body 
decays as:
\begin{eqnarray}
 BR(\Dm_L \to \tilde{\ell}_{iL}^-\ell_i^-)=\frac{1}{3},&&i=e,\mu,\tau,
~~~~m_{\tilde l_{il}}<M_{\tilde\Delta_L^{--}} \nonumber\\
 BR(\tilde{\ell}_{iL}^- \to \ell_i^- \N0_1)= 1,&&i=e,\mu,\tau \nonumber\\
 BR(\Dm_L \to \ell_i^- \ell_i^-  \N0_1)= \frac{1}{3},&&i=e,\mu,\tau,
~~~~m_{\tilde l_{il}}>M_{\tilde\Delta_L^{--}} \nonumber
\end{eqnarray}
where only the 3-body decay is allowed when $m_{\tilde{\ell}_{iL}}>M_{\Dm_L}$.
To define our notations, we write {\it signal-1} to correspond to the 
2-body decay of $\Dm_L$ and write {\it signal-2} to correspond to the
3-body decay of $\Dm_L$. 
We use {\it sample point} {\bf A} for the machine with $\sqrt{s}=500$ GeV as its
center-of-mass energy and {\it sample point} {\bf B} for the  $\sqrt{s}=1$ TeV
option. We choose $M_{\Dm_L}=300$ GeV and $m_{\tilde{\ell}_{iL}}=150$ GeV 
for {\it signal-1}, while for {\it signal-2}, $m_{\tilde{\ell}_{iL}}=400$ GeV 
to study the signal at the $\sqrt{s}=500$ GeV machine. The corresponding choice
for the analysis at the $\sqrt{s}=1$ TeV option is  $M_{\Dm_L}=500$ GeV,
$m_{\tilde{\ell}_{iL}}=250$ GeV for {\it signal-1}, while for {\it signal-2}, 
$m_{\tilde{\ell}_{iL}}=550$ GeV.  

The background for the signal (ii) $\mu^-\mu^- E\slash$ is a six-body
final state and would be very small. 
The cuts, which gave the SM background $\sigma (e^- e^- E\slash)$=
537.7 fb/1.13 pb,
give cross sections: 
$\sigma (\mu^-\mu^- E\slash) \sim {\mathcal O(10^{-1})/O(1)}~fb$ for, respectively, the
$\sqrt{s}=500/1000$ GeV machine.
However the effect of cuts are not severe for the $\mu^-\mu^- E\slash$ final 
states as compared to $e^- e^- E\slash$ final states, where the cross section is 
reduced by more than $\sim 80-90\%$ of the original one, while
the cross section in this case reduces by about 30\% for the more stringent 
cuts listed in Table~\ref{rates}. Hence we have chosen to use the same set of
cuts for both final states.

We present our results for the set of cuts which reduces the background while 
keeping the signal relatively large. Our choice of cuts for the 
signal-background analysis is as follows:
\begin{itemize}
\item We demand that the electrons respect a minimum rapidity cut of 
$|\eta_{e}|<1.5$ which is the most effective cut to reduce the large continuum
background. The signal is peaked at $\eta_e=0$, while the SM background is
peaked for $|\eta_e|> 1.5$, which is predominantly due to the strong t-channel
photon contribution in the $e^- e^- E\slash$ final states.
\item The final state electrons must carry a minimum energy,  $E_e>5$ GeV.
\item We also demand a large cut on the missing energy because of the 
massive LSP in the final state for the signal, $E\slash > \sqrt{s}/2$ GeV.
\item To ensure proper resolution between the final state electrons we demand
that they are well separated in space and satisfy $\D R_{ee}>0.2$.
\end{itemize}
Using the above cuts, we can see from Table~\ref{rates} that the large SM 
background which previously overwhelmed the signal for the most conservative 
cuts, is effectively suppressed without loosing out much on the signal events. 

\begin{figure}[htb]
\begin{center}
\includegraphics[height=2.8in,width=2.8in]{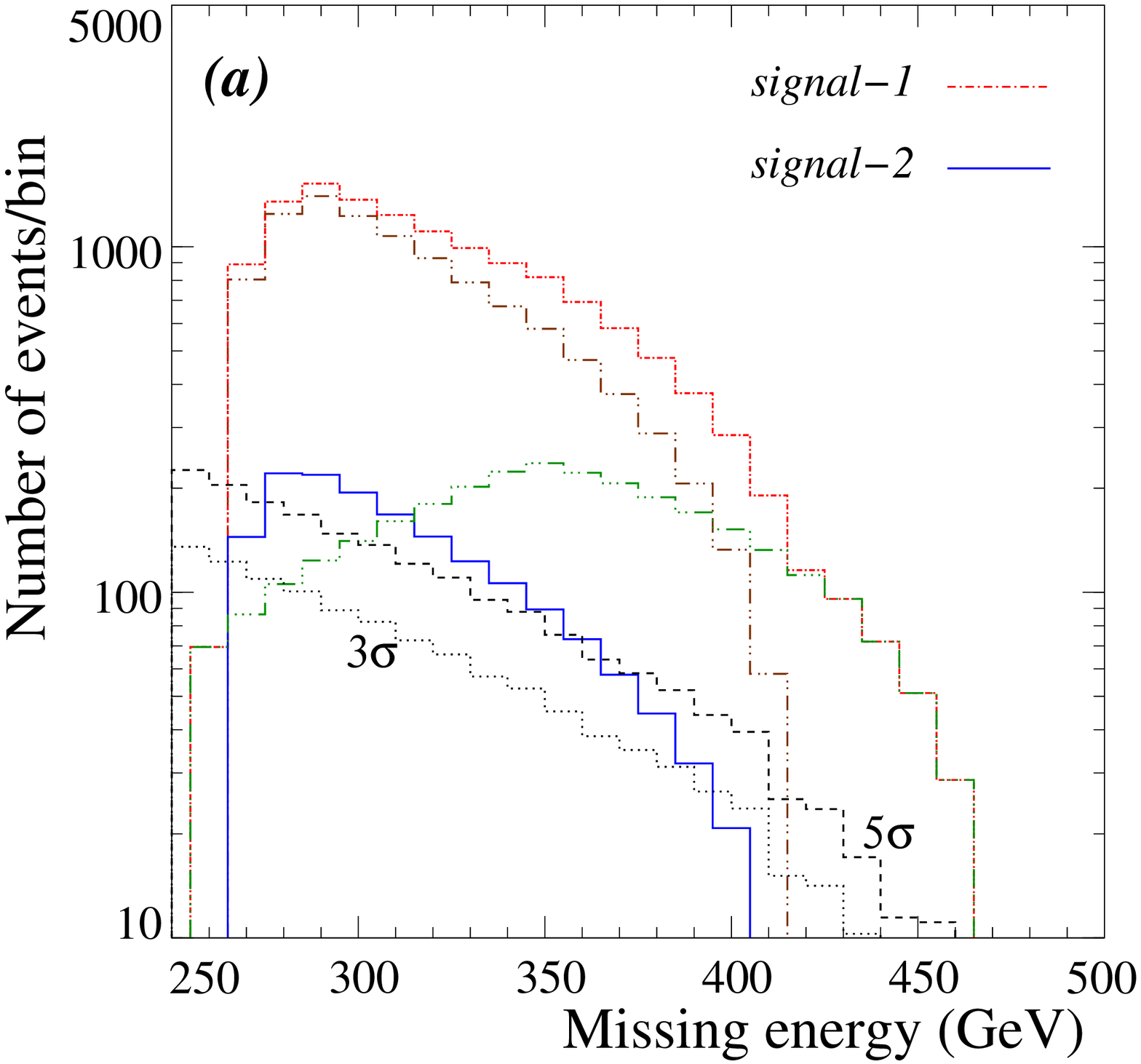} 
\includegraphics[height=2.8in,width=2.8in]{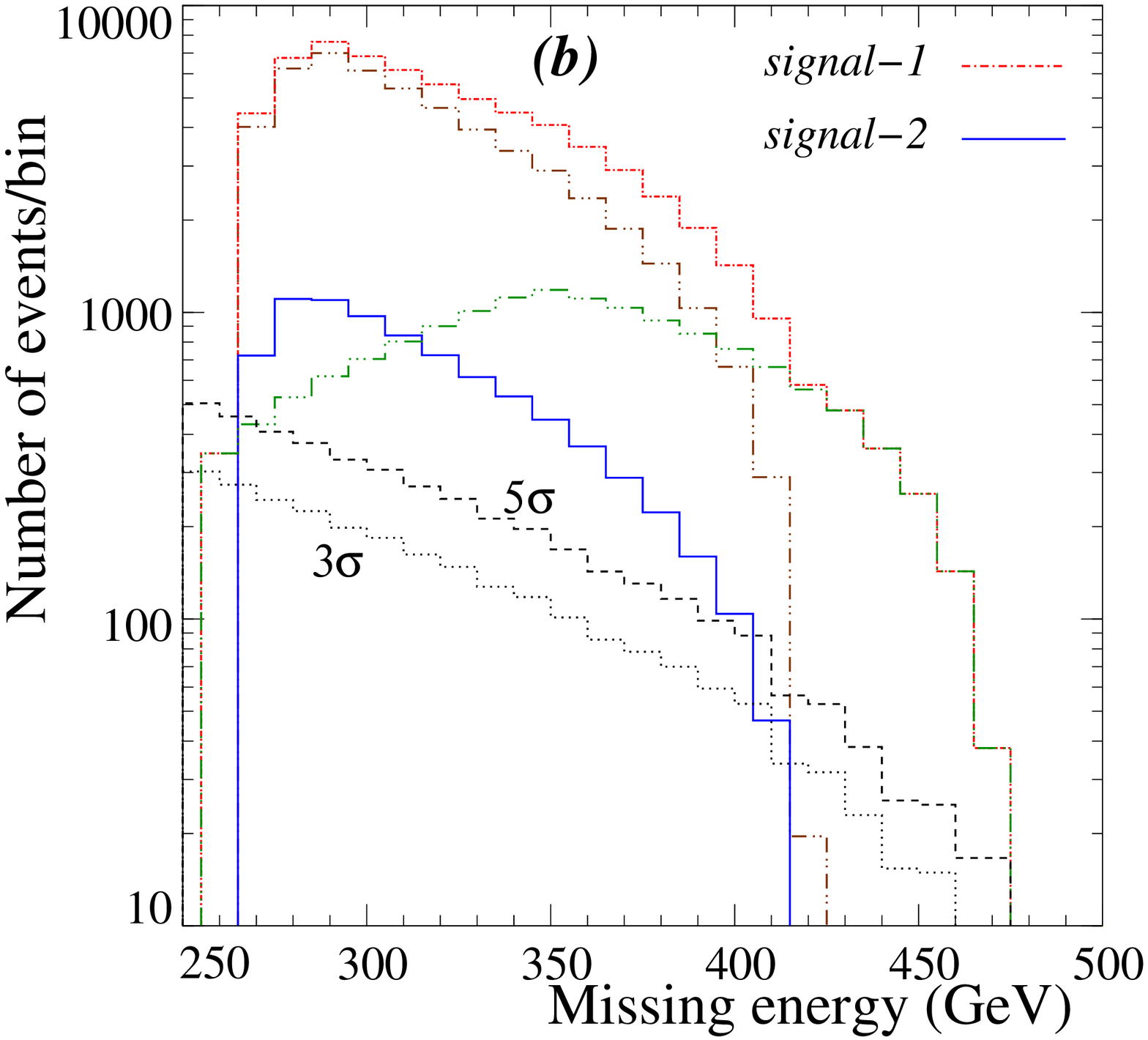}
\caption{\sl\small Binwise distribution of missing energy for both 
{\it signal-1} and {\it signal-2} for the final states $e^-e^-E\slash$
($\sqrt{s}=500$ GeV). 
The broken ($-\cdot\cdot-$) brown lines correspond to the signal 
through $\Dm_L$ production while the broken ($-\cdots-$) green lines 
correspond to the signal events from $\tilde{e}^-_L$-pair production.
Also shown in dashed and dotted dark 
lines are the $5\sigma$ and $3\sigma$ fluctuations in the SM background. 
Each binsize is 10 GeV. (a) ${\cal L}= 100 fb^{-1}$, 
(b) ${\cal L}=500 fb^{-1}$.}
\label{energymiss1}
\end{center}
\end{figure}

\begin{figure}[htb]
\begin{center}
\includegraphics[height=2.8in,width=2.8in]{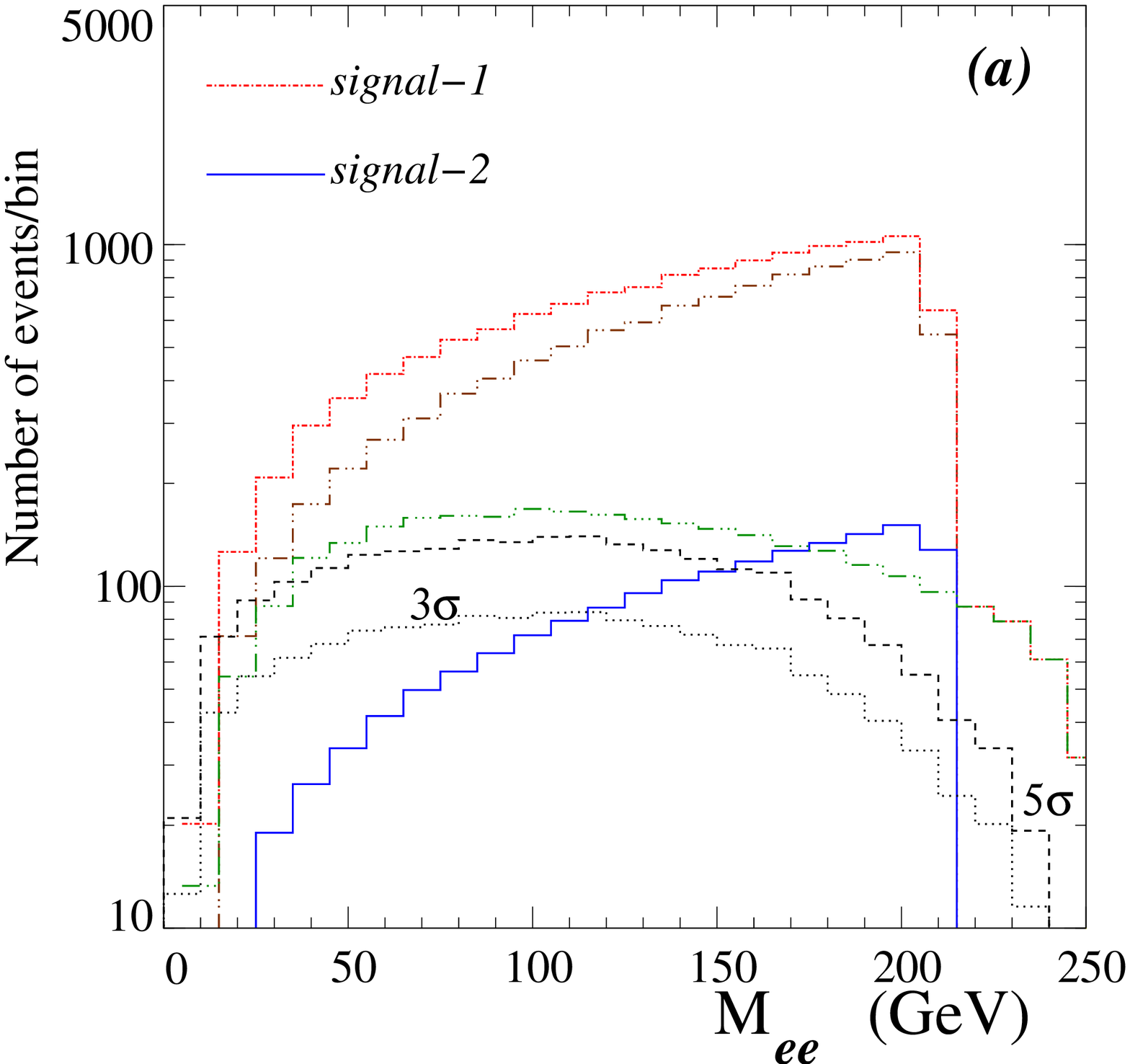} 
\includegraphics[height=2.8in,width=2.8in]{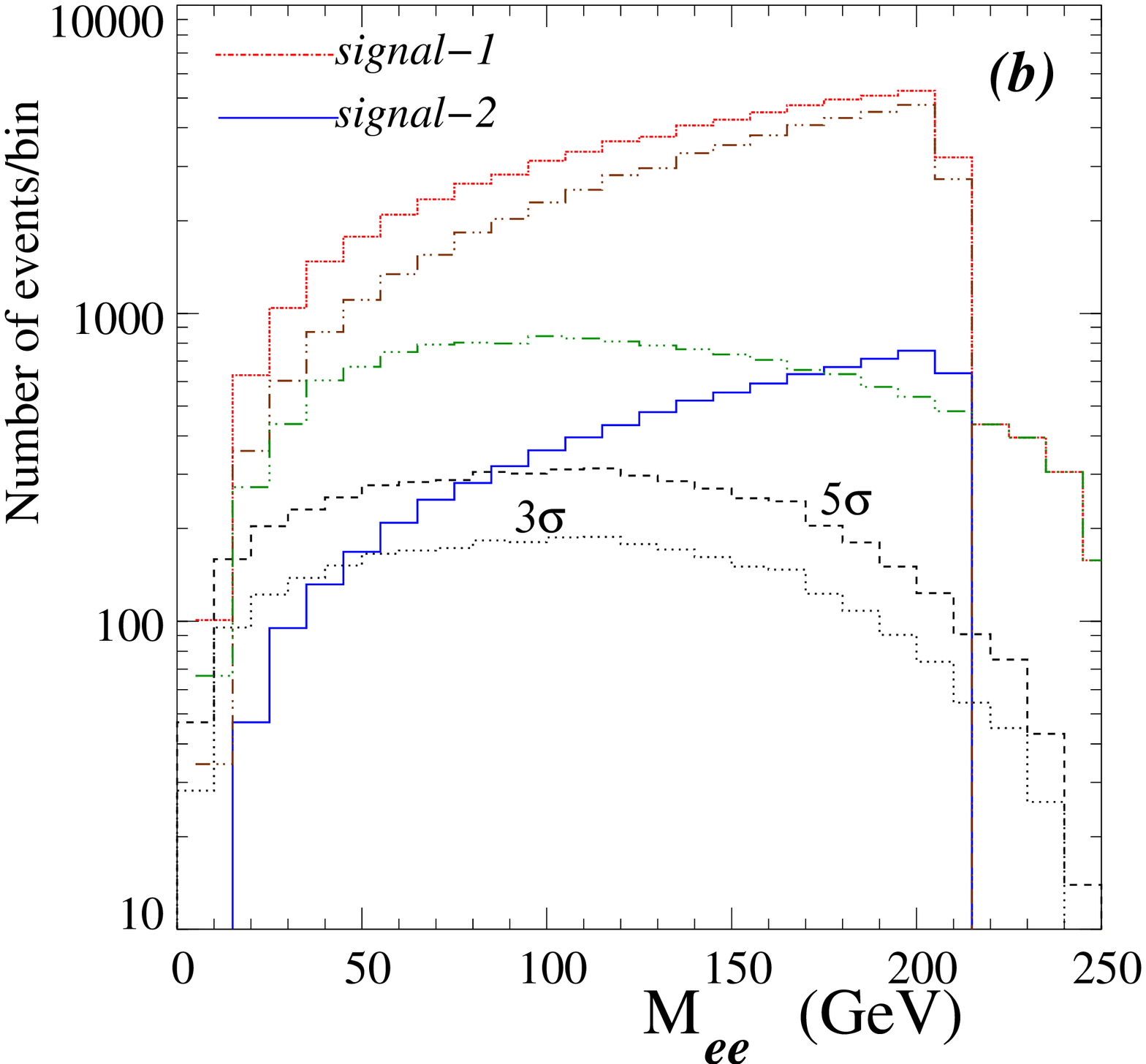}
\caption{\sl\small Binwise distribution of the invariant mass of the visible 
particles in the final state $e^-e^-E\slash$ for both {\it signal-1} and
{\it signal-2} ($\sqrt{s}=500$ GeV). 
The broken ($-\cdot\cdot-$) brown lines correspond to the signal 
through $\Dm_L$ production while the broken ($-\cdots-$) green lines 
correspond to the signal events from $\tilde{e}^-_L$-pair production.
The background follows the notation of Fig~\ref{energymiss1}. 
Each binsize is 10 GeV.(a) ${\cal L}=100 fb^{-1}$, (b) ${\cal L}=500 fb^{-1}$.}
\label{invmass1}
\end{center}
\end{figure}

\begin{figure}[htb]
\begin{center}
\includegraphics[height=2.8in,width=2.8in]{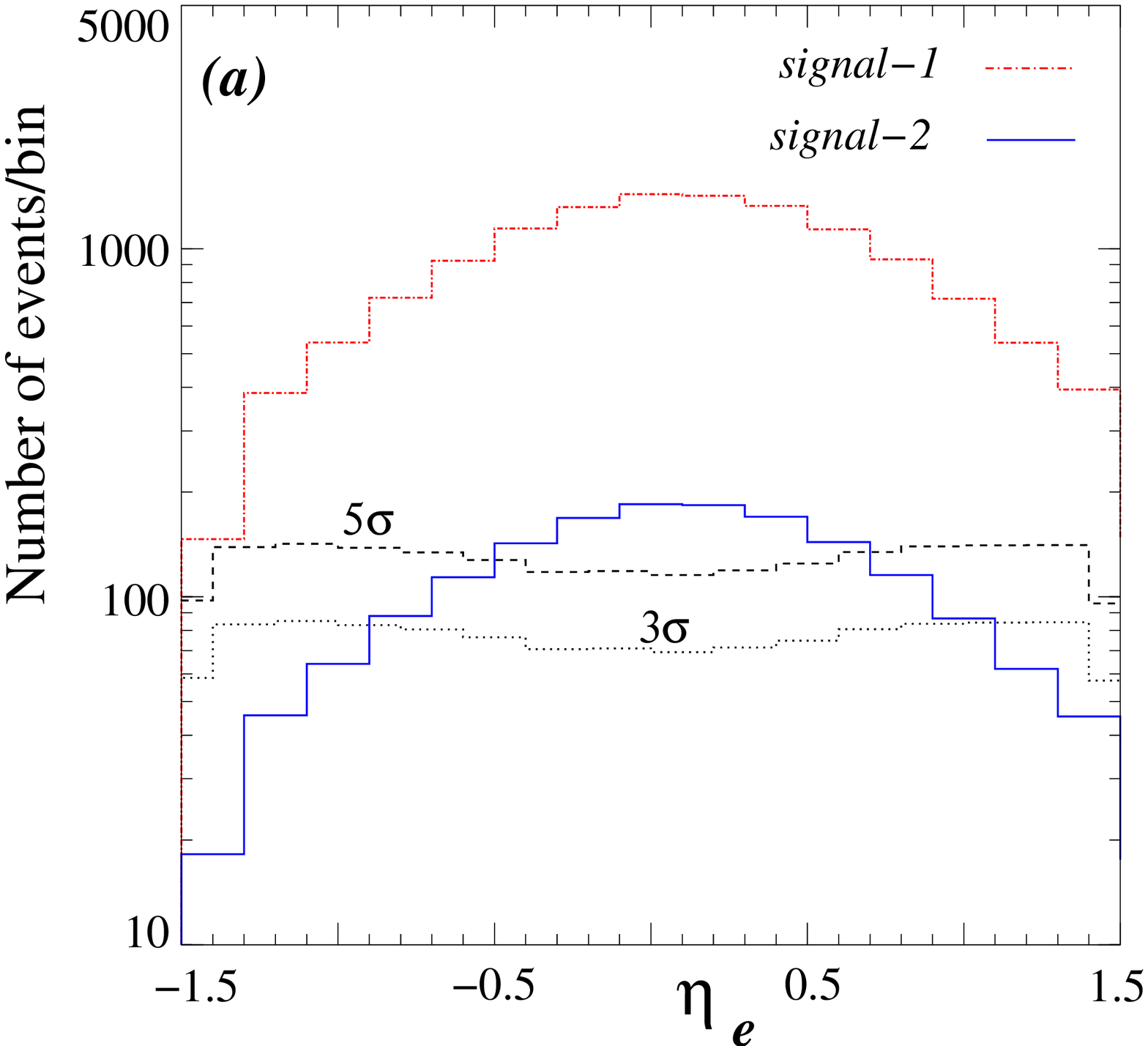} 
\includegraphics[height=2.8in,width=2.8in]{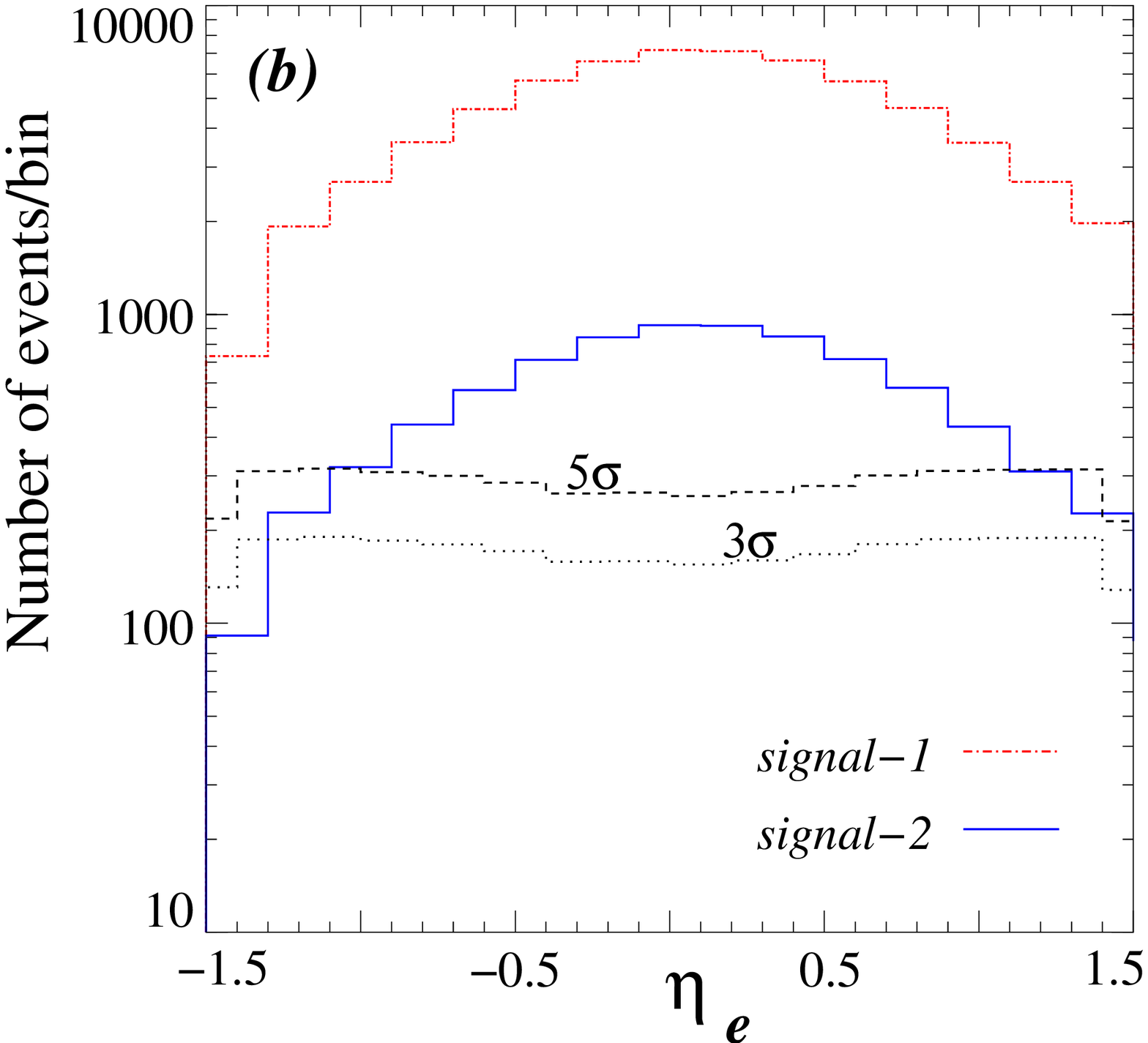}
\caption{\sl\small Binwise distribution of the rapidity of the $e^-$ in the final state $e^-e^-E\slash$ for both {\it signal-1} and
{\it signal-2} ($\sqrt{s}=1$ TeV).
The background follows the notation of Fig~\ref{energymiss1}. Each binsize is 
0.2 radians. (a) ${\cal L}=100 fb^{-1}$, (b) ${\cal L}=500 fb^{-1}$.}
\label{eta1}
\end{center}
\end{figure}

\begin{figure}[htb]
\begin{center}
\includegraphics[height=2.8in,width=2.8in]{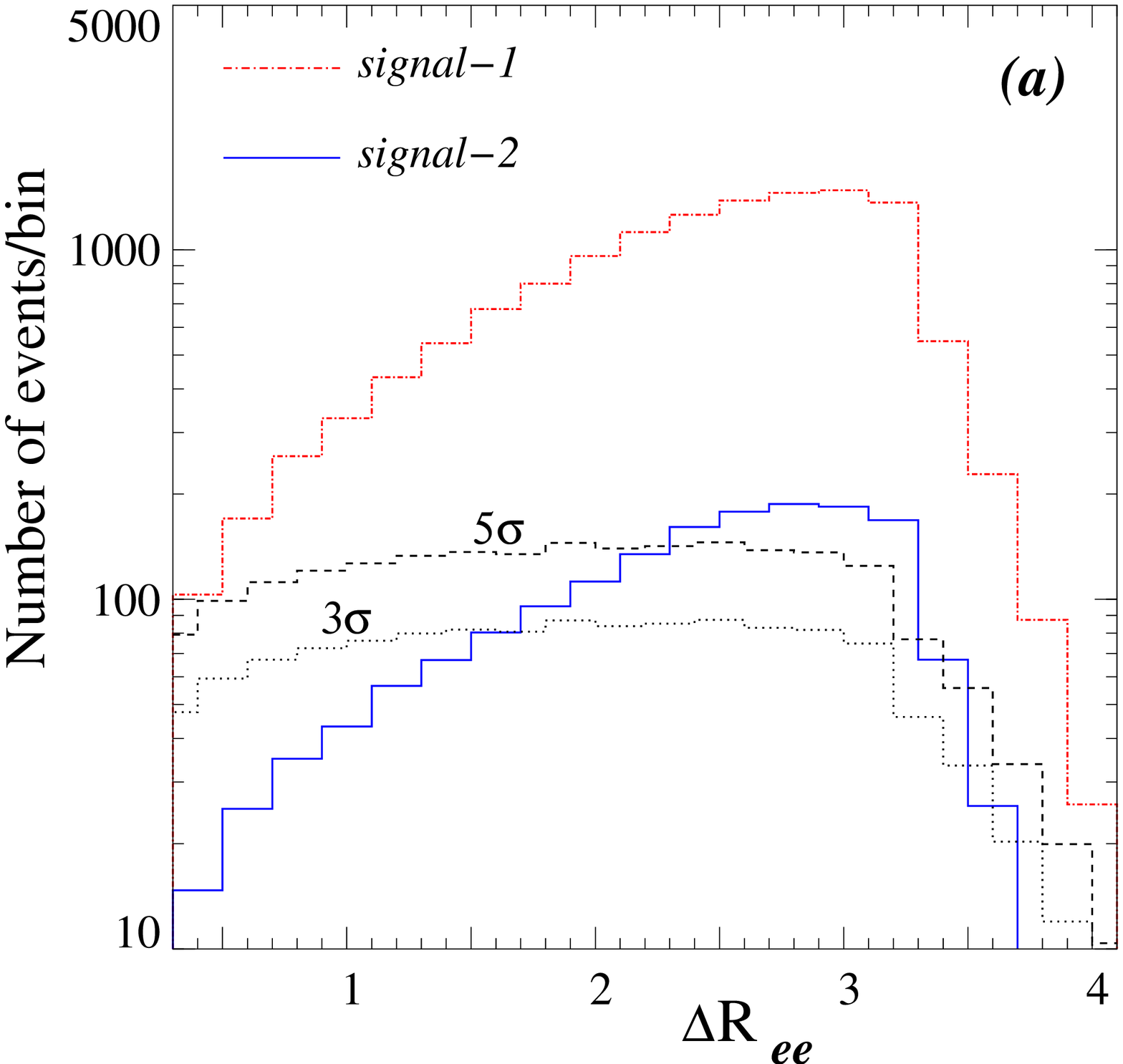} 
\includegraphics[height=2.8in,width=2.8in]{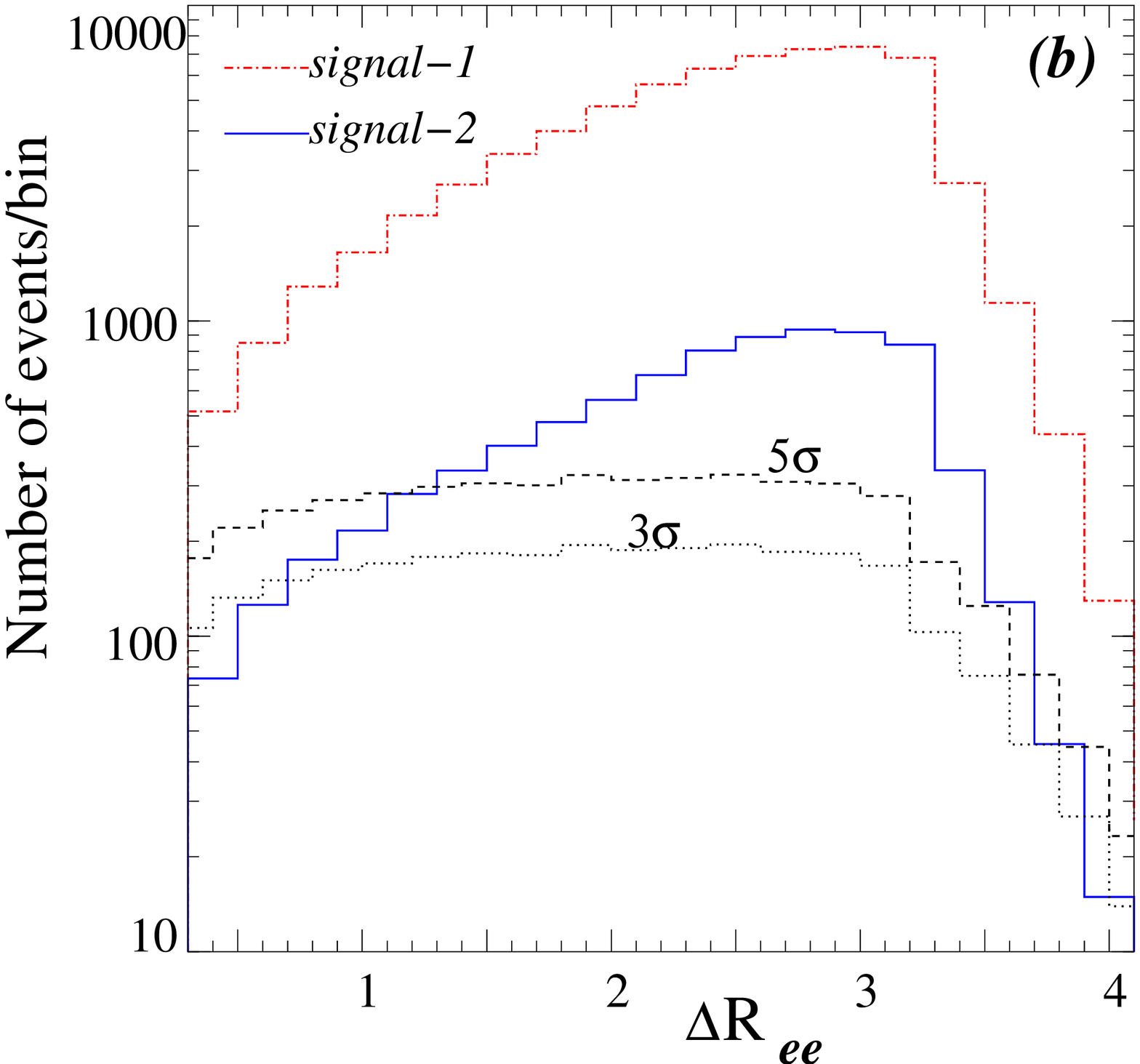}
\caption{\sl\small Binwise distribution of the $\D R$ between the visible
particles in the final state $e^-e^-E\slash$ for both {\it signal-1} 
and {\it signal-2} ($\sqrt{s}=500$ GeV).
The background follows the notation of Fig~\ref{energymiss1}. 
Each binsize is 0.2. (a) ${\cal L}=100 fb^{-1}$, (b) ${\cal L}=500 fb^{-1}$.}
\label{deltaR1}
\end{center}
\end{figure}

\begin{figure}[htb]
\begin{center}
\includegraphics[height=2.8in,width=2.8in]{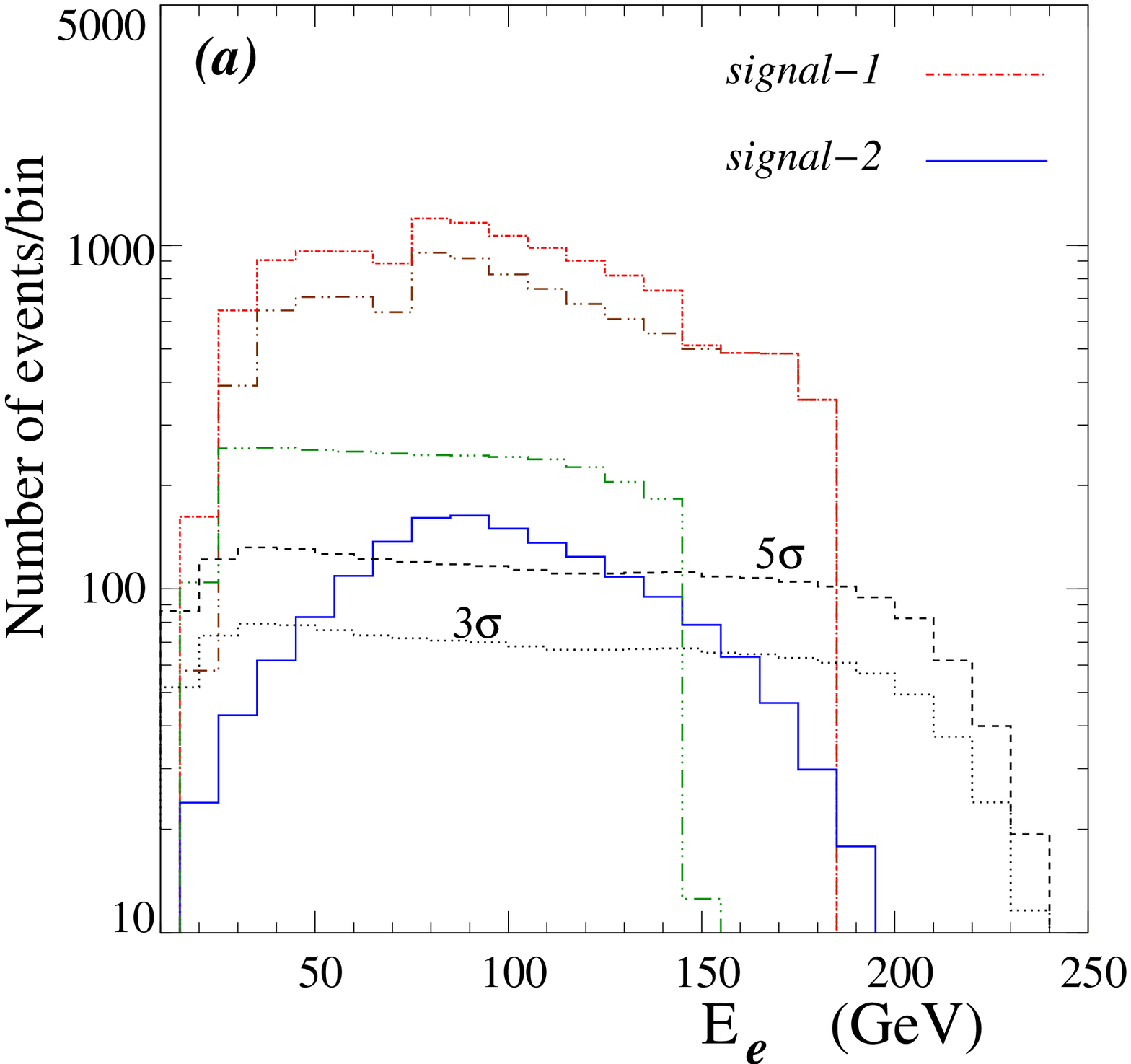} 
\includegraphics[height=2.8in,width=2.8in]{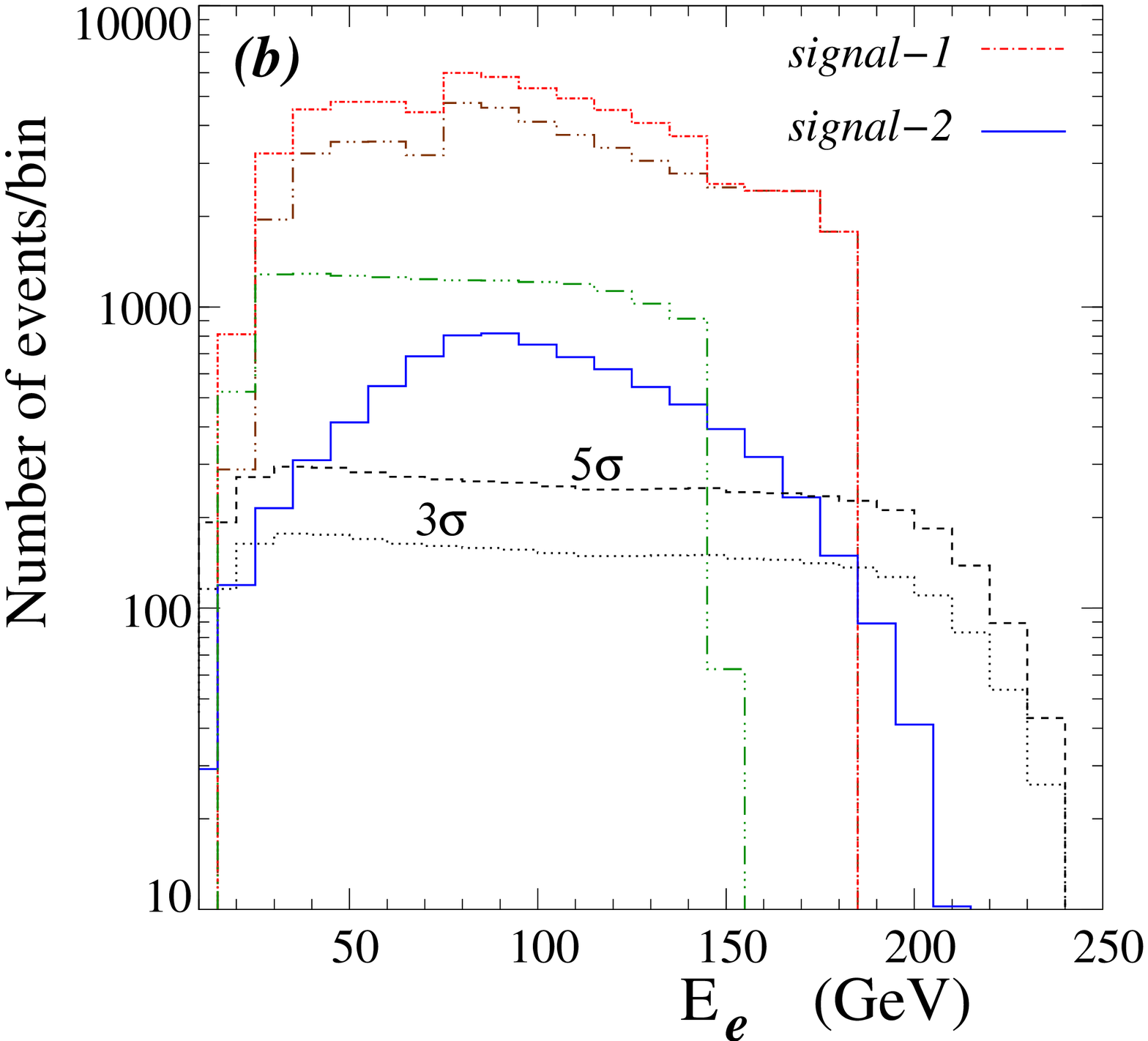}
\caption{\sl\small Binwise distribution of the energy of $e^-$
in the final state $e^-e^-E\slash$ for both {\it signal-1} and
{\it signal-2} ($\sqrt{s}=500$ GeV).
The broken ($-\cdot\cdot-$) brown lines correspond to the signal 
through $\Dm_L$ production while the broken ($-\cdots-$) green lines 
correspond to the signal events from $\tilde{e}^-_L$-pair production.
The background follows the notation of Fig~\ref{energymiss1}. Each binsize is 
 10 GeV. (a) ${\cal L}=100 fb^{-1}$, (b) ${\cal L}=500 fb^{-1}$.}
\label{energydist1}
\end{center}
\end{figure}

We present the different kinematic distributions in 
figures~\ref{energymiss1}, \ref{invmass1}, \ref{eta1}, \ref{deltaR1} and 
\ref{energydist1} for the final states $e^-e^-E\slash$  
for the 500 GeV machine. We must point out that the SM background for the
other final state $\mu^-\mu^-E\slash$ at the 500 GeV machine turns out to be
completely insignificant and the signal would be background free. In 
fact the signal would also act as an obvious hint for scenarios with doubly 
charged higgsinos and provide strong distinguishing features from the typical 
MSSM signal coming from selectron pair 
production ($e^-e^-\to\tilde{e}^-_L\tilde{e}^-_L$), which produces an 
identical final state $e^-e^-E\slash$ but gives no contribution to 
the $\mu^-\mu^-E\slash$ final states. 
Having said this, we must note that for the $e^-e^-E\slash$ final states,
we have additional events coming from $e^-e^-\to\tilde{e}^-_L\tilde{e}^-_L$ in 
this model too. Looking at our mass spectrum, this production mode will only 
contribute to {\it signal-1}. We estimate this to be $29.72~fb$ at the 
500 GeV machine for the polarization choice (-1,-1) and the same set of cuts. 
We include this contribution in {\it signal-1} for all the figures 
(\ref{energymiss1}--\ref{energydist1}), and
the integrated luminosity taken in {\it (a)} and {\it (b)} is 
100 and 500 $fb^{-1}$ respectively. We have also shown the independent 
contributions for {\it signal-1} coming from $\Dm_L$ and $\tilde{e}^-_L$-pair 
production in Figs.~\ref{energymiss1}, \ref{invmass1} and \ref{energydist1} to
highlight the difference in the two cases. 
As mentioned earlier, the doubly charged 
higgsino mass is taken to be 300 GeV, while the selectron masses are 150 GeV 
and 400 GeV for {\it signal-1} and {\it signal-2} respectively.

In Fig.~\ref{energymiss1} we plot the binwise distribution of missing energy for 
both {\it signal-1} and {\it signal-2}, represented by colored histograms, 
and compare it with the expected statistical (Gaussian) fluctuations at the 3 
and 5 standard deviations of the SM background. As expected, the signal would 
be associated with a large missing energy due to the presence of LSP in the 
final state, and this also allows to set a strong cut on the missing energy. In 
Fig.~\ref{energymiss1}(a) we show the distribution for an integrated luminosity
of 100 $fb^{-1}$ and find that {\it signal-1} stands out quite distinctly 
against a $5\sigma$ fluctuation in the SM background. The fact that the signal 
is actually superimposed over the tail end of the much wider missing energy 
distribution of the SM, carried by the neutrinos, makes {\it signal-1} 
stand out against such large fluctuations in the SM background. We show 
the same distribution in Fig.~\ref{energymiss1}(b) with an integrated 
luminosity of 500 $fb^{-1}$, where both {\it signal-1} and {\it signal-2}
stand out against a $5\sigma$ fluctuations in the background. 

In Fig.~\ref{invmass1} we show the invariant mass distribution for the pair of 
visible particles in the final state. As both the visible leptons come from the 
(cascade) decay of $\Dm_L$, the distribution for the signal would  
depend strongly on the masses of $\Dm_L$ and $\N0_1$'s in the final 
state, and the distribution exhibits a sharp kinematic edge which highlights 
this dependence. The location of this kinematic edge can be well approximated 
by the formula
$$
M_{ee}^{edge} \approx \sqrt{M_{\Dm_L}^2 + M_{\N0_1}^2 - \sqrt{s}M_{\N0_1}}
$$
where we have assumed that the final neutralino coming from the cascade 
decay of $\Dm_L$ carries a negligible 3-momentum, as its mass is nearly a third 
of the mass of $\Dm_L$. This is in sharp contrast to 
the SM distribution, where the background has a more uniform invariant mass 
distribution. In fact, the distribution arising from the independent 
contribution of $\tilde{e}^-_L$-pair production shown in broken ($-\cdots-$)
green line has a uniform distribution like the SM background, but is 
overwhelmed by the events coming from the $\Dm_L$ production. 
The luminosity choices are as  for the plots in Fig.~\ref{invmass1}(a) and (b). 

In Fig.~\ref{eta1} and Fig.~\ref{deltaR1}, we give the rapidity distribution 
for the final state leptons and the $\D R$ distribution between the pair of 
visible particles in the final state respectively. The rapidity distribution 
clearly suggests that 
the leptons in the signal distribution emerge back to back, since the 
distribution peaks at $\eta=0$. This is because the leptons are decay products
of a massive $\Dm$ which has low boost, and hence it is more likely to give off 
back to back lepton-slepton in {\it signal-1}, while the other lepton coming 
\begin{figure}[htb]
\begin{center}
\includegraphics[height=2.3in,width=2.3in]{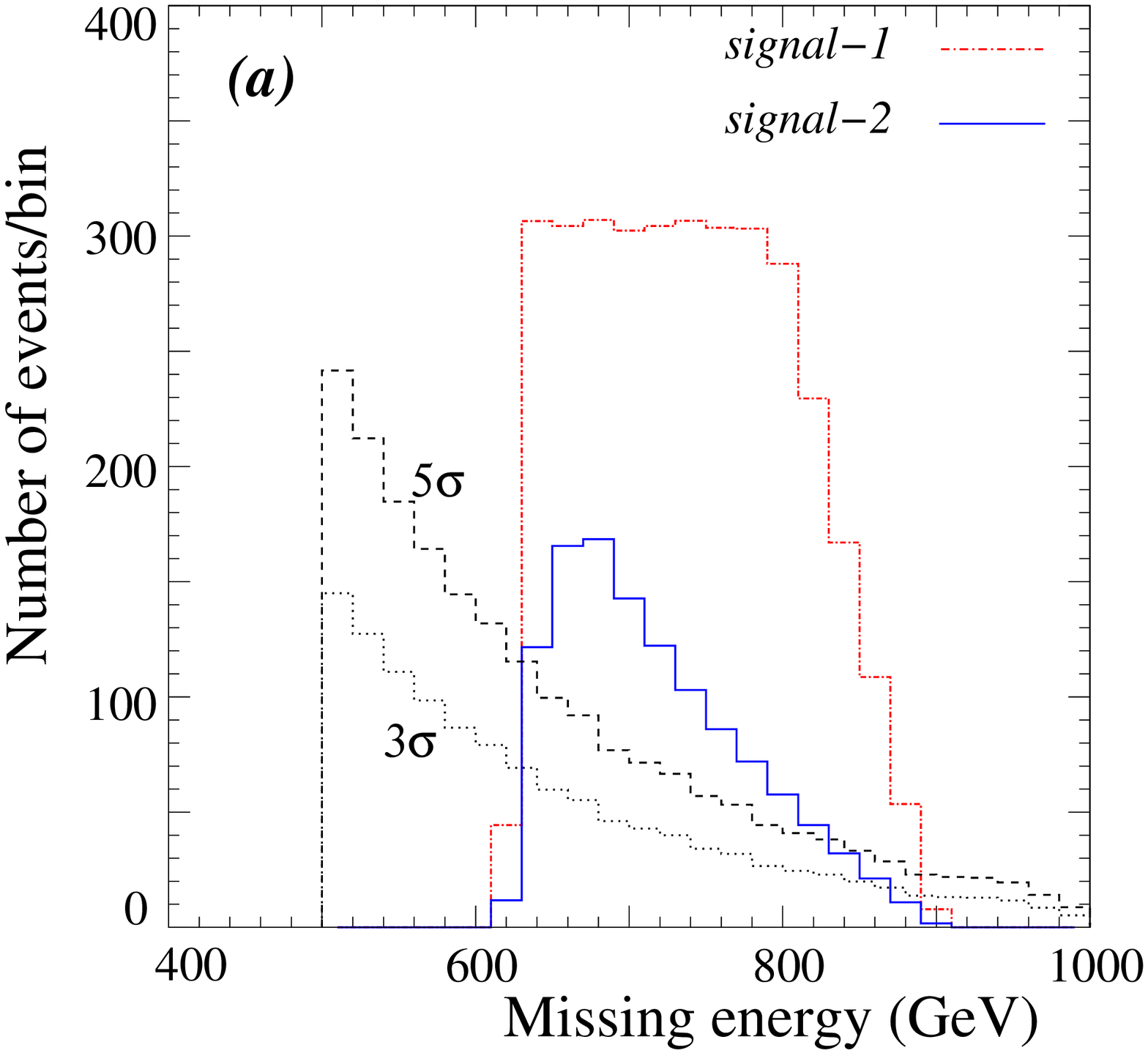}
\includegraphics[height=2.3in,width=2.3in]{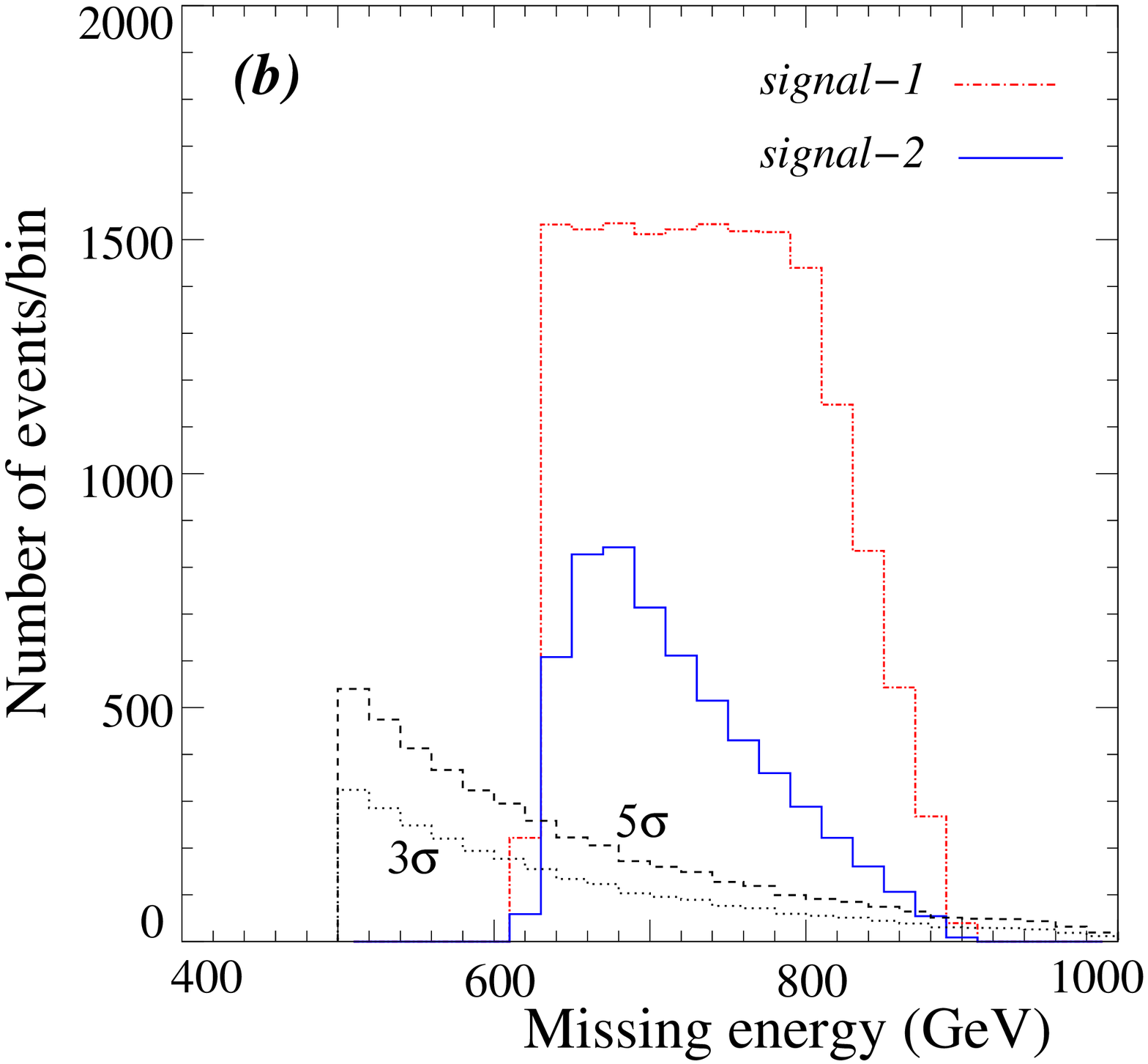}
\includegraphics[height=2.3in,width=2.3in]{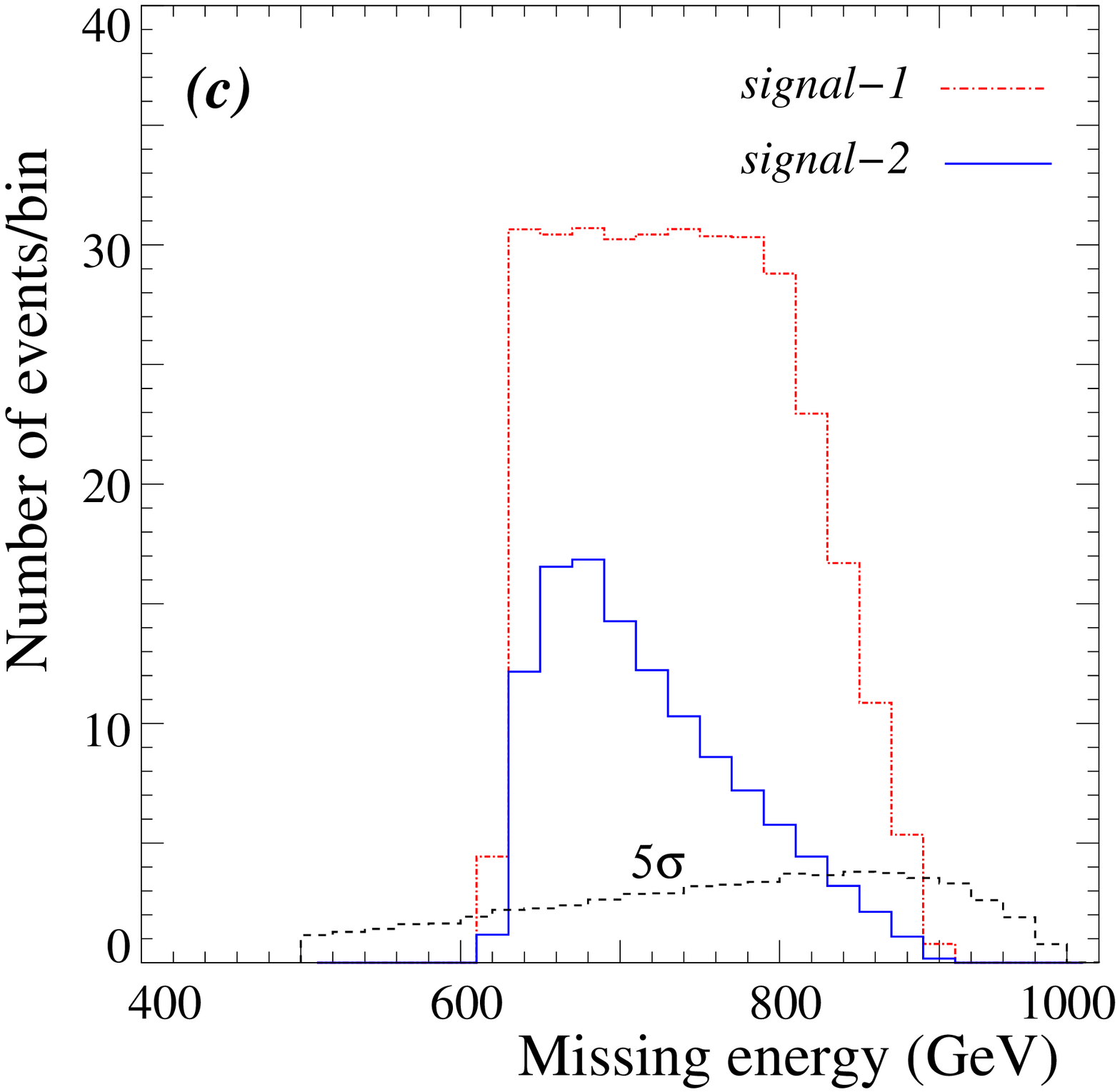}
\caption{\sl\small Binwise distribution of missing energy for both 
{\it signal-1} and {\it signal-2} ($\sqrt{s}=1$ TeV). 
The background follows the notation of Fig~\ref{energymiss1}. Each binsize is 
20 GeV. (a) ${\cal L}=100 fb^{-1}$ (final state $e^-e^-E\slash$), 
(b) ${\cal L}=500 fb^{-1}$  (final state $e^-e^-E\slash$), 
(c) ${\cal L}=10 fb^{-1}$  (final state $\mu^-\mu^-E\slash$).
Here $M_{\tilde\Delta_L^{--}}=$500 GeV. For {\it signal-1} 
$m_{\tilde e_L}=$250 GeV  and for {\it signal-2 $m_{\tilde e_L}=$500 GeV.}
}
\label{energymiss2}
\end{center}
\end{figure}
from the heavy slepton is most likely to carry the maximum boost of the 
slepton as it is produced against a massive particle $(\N0_1)$ with less
boost. The distribution for {\it signal-2} shows a similar behavior 
\begin{figure}[htb]
\begin{center}
\includegraphics[height=2.3in,width=2.3in]{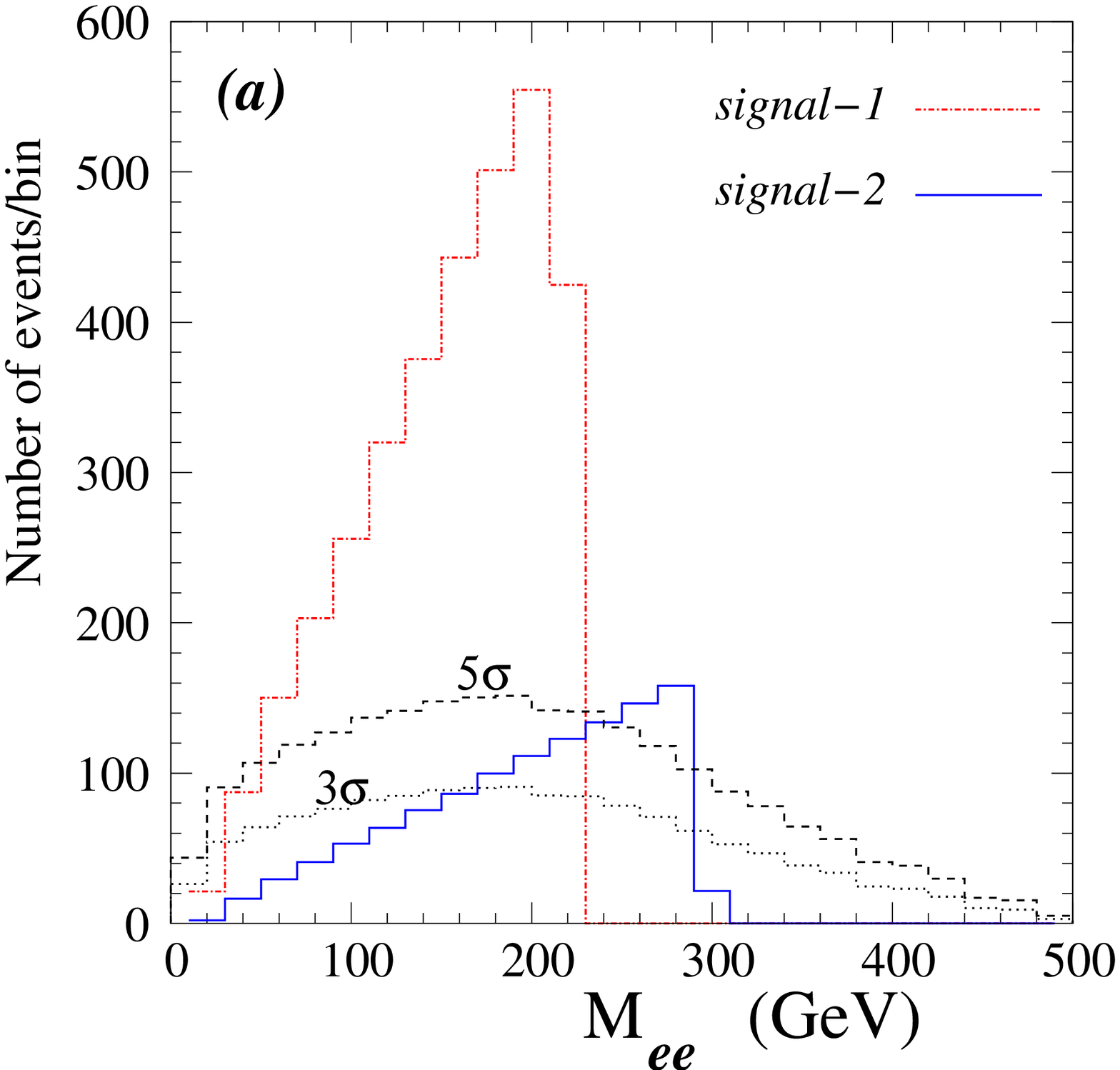}
\includegraphics[height=2.3in,width=2.3in]{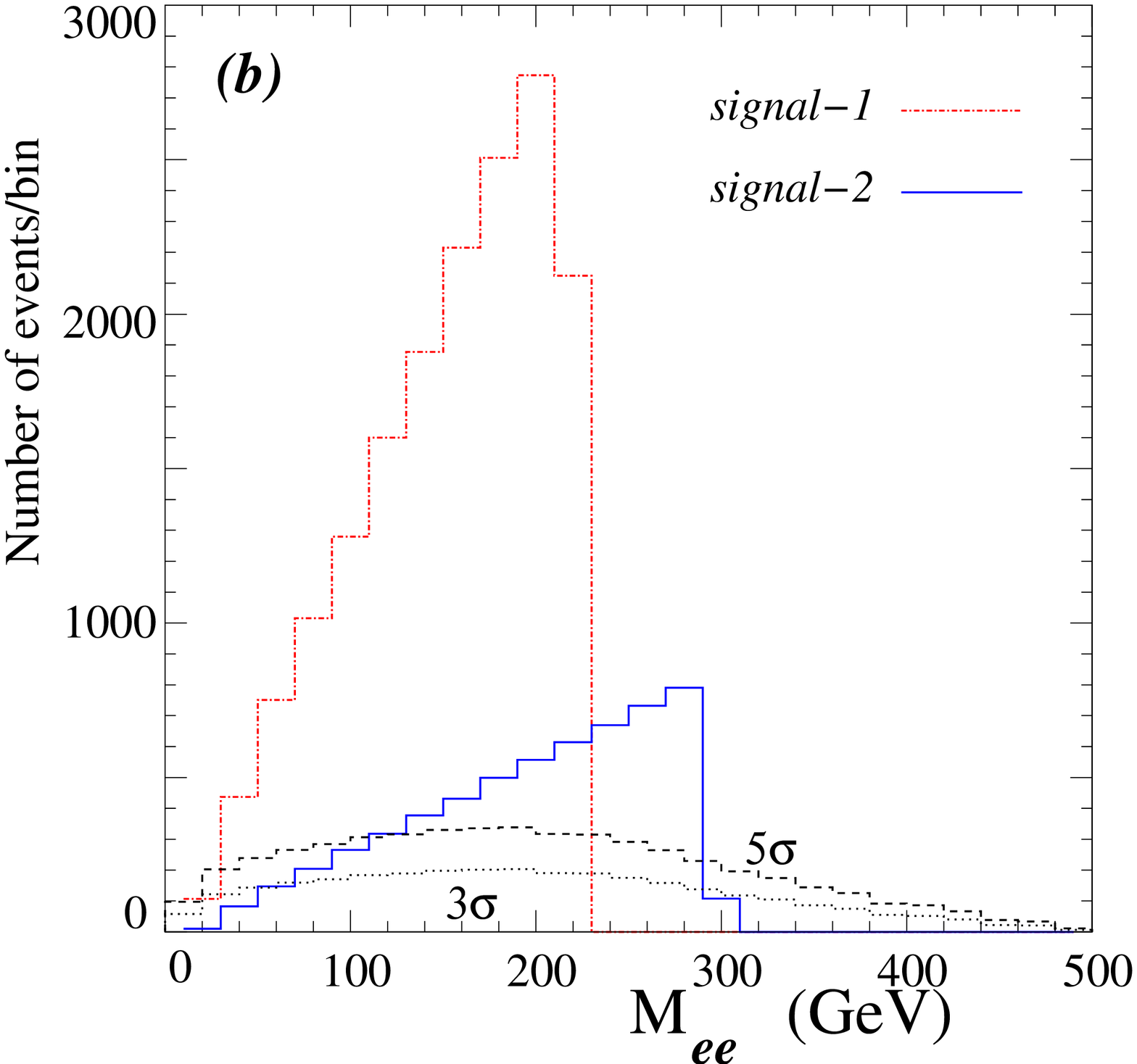}
\includegraphics[height=2.3in,width=2.3in]{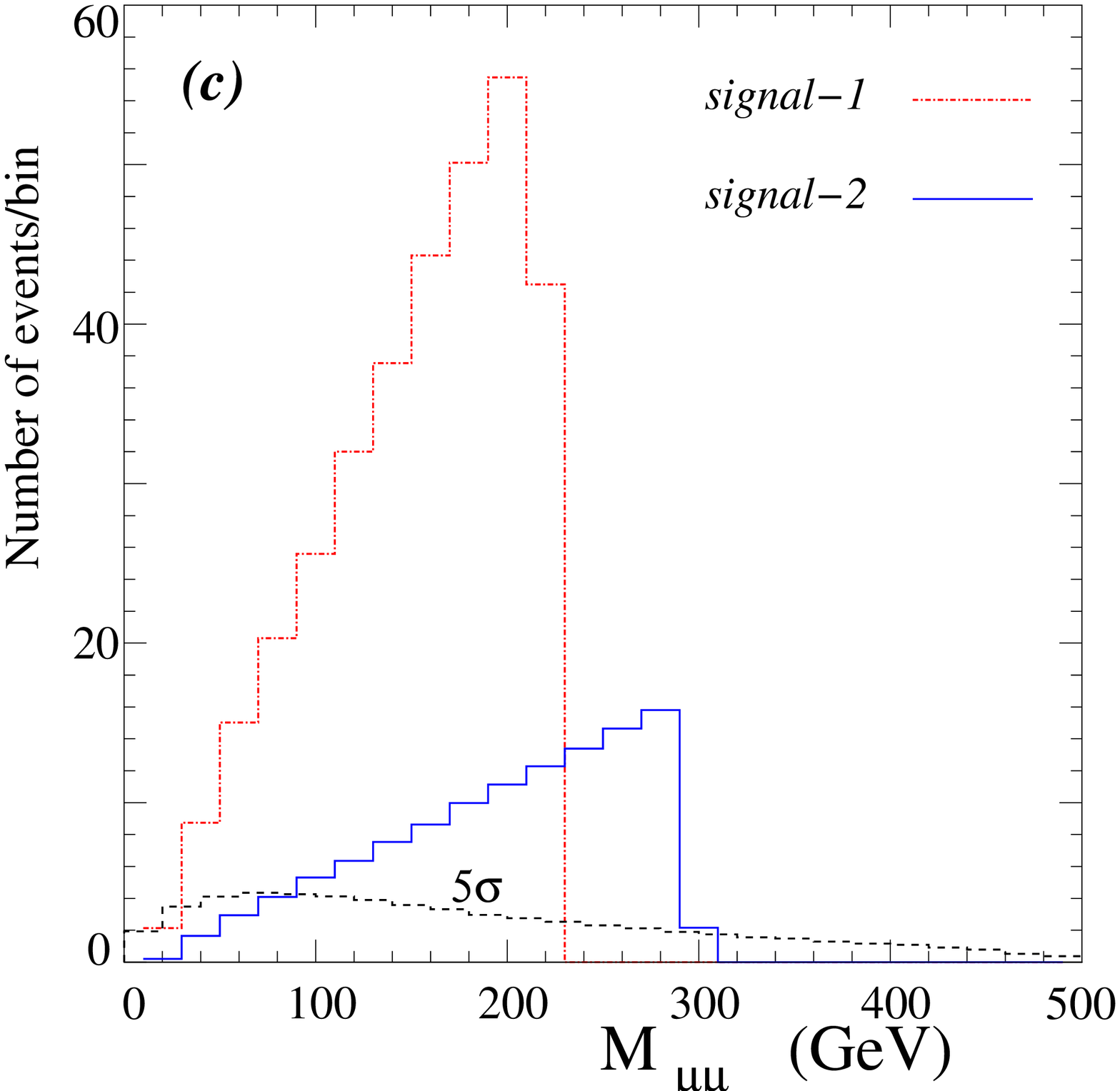}
\caption{\sl\small Binwise distribution of the invariant mass of the visible 
particles in the final state for both {\it signal-1} and {\it signal-2}. 
Each binsize is 20 GeV ($\sqrt{s}=1$ TeV). 
The background follows the notation of Fig~\ref{energymiss1}.
(a) ${\cal L}=100 fb^{-1}$ (final state $e^-e^-E\slash$),
(b) ${\cal L}=500 fb^{-1}$  (final state $e^-e^-E\slash$), 
(c) ${\cal L}=10 fb^{-1}$  (final state $\mu^-\mu^-E\slash$). 
Here $\sqrt{s}=1$ TeV and $M_{\tilde\Delta_L^{--}}=$500 GeV.
For {\it signal-1} $m_{\tilde e_L}=$250 GeV  and
for {\it signal-2 $m_{\tilde e_L}=$500 GeV.}}

\label{invmass2}
\end{center}
\end{figure}
because the 2-lepton system recoils against a massive LSP which would have a 
minimum boost. For the background, the distribution is mostly peaked at 
large values of $|\eta|$ due to the strong $t$-channel contribution from 
the photon exchange, as discussed earlier. This actually helps  
suppress the huge continuum background in the case of $e^-e^-E\slash$ 
final states, as shown in Table~\ref{rates}.
The distribution in $\D R_{ee}$ also shows a clear difference in behavior
from the SM background, which can again be attributed to the fact that the 
angular distributions for the final state charged leptons in the signal events 
are markedly different from those of the SM events, and if one considers 
$$|\D\eta|=|\eta_{e^-}^1 - \eta_{e^-}^2|$$ the distribution actually shows a
peak for the signal at $|\D\eta|=0$, while for 
the SM background the signal peaks beyond $|\D\eta|>1.2$. 

Finally we present the energy distribution of the charged leptons for the 
two different final states in Fig.~\ref{energydist1}. The signal is again seen 
to stand out against the $5\sigma$ fluctuations in the SM background for both 
{\it signal-1} and {\it signal-2} for the higher integrated luminosity of 
500 $fb^{-1}$ in Fig.~\ref{energydist1}(b). The more interesting feature is seen
for {\it signal-1}, where the energy distributions for the final two charged 
leptons will in principle be different. Assuming that it would be difficult to 
distinguish between the two leptons, we show the distribution by taking the 
average of both the leptons energy distributions. However, in the case of 
\begin{figure}[htb]
\begin{center}
\includegraphics[height=2.3in,width=2.3in]{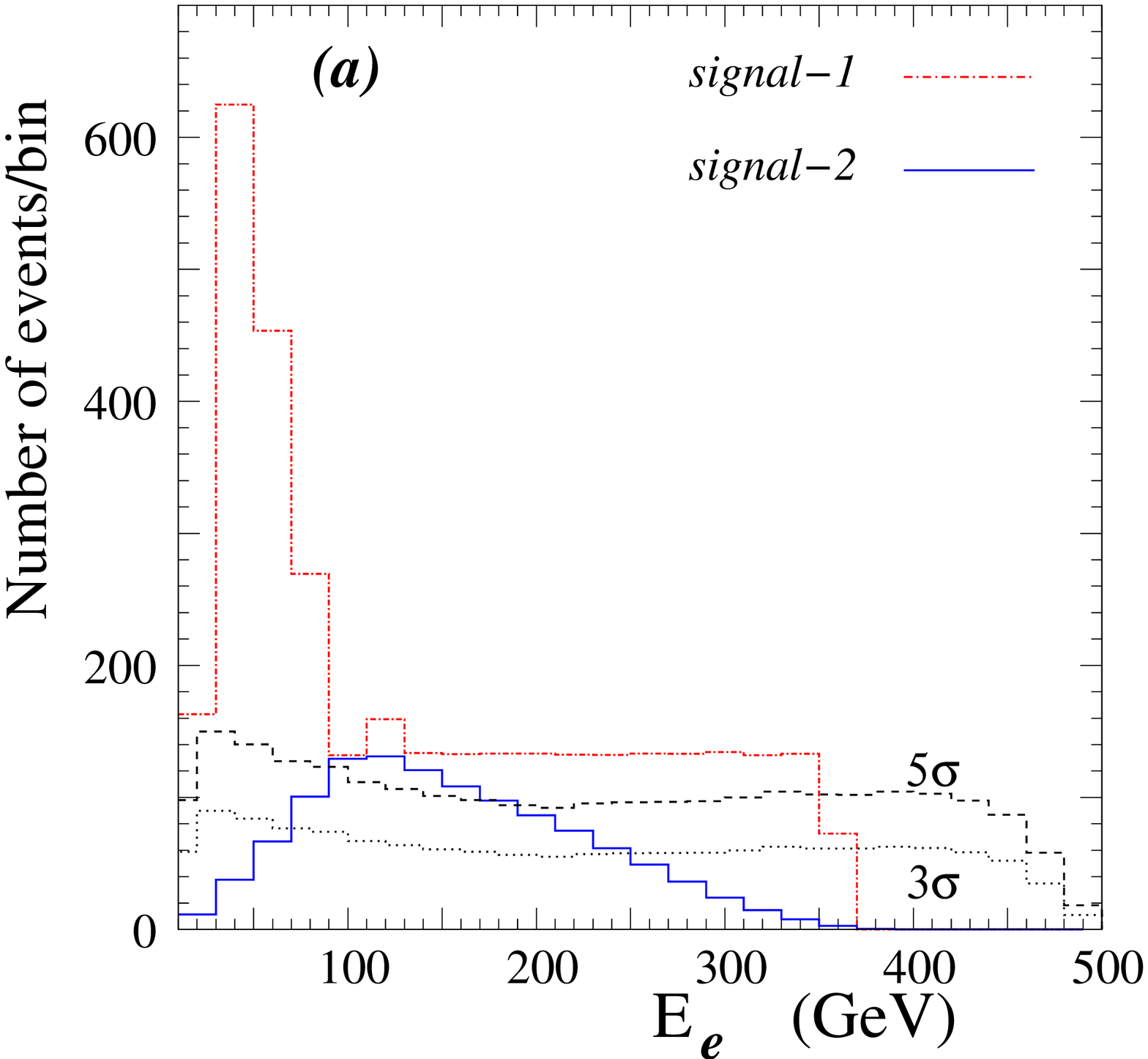}
\includegraphics[height=2.3in,width=2.3in]{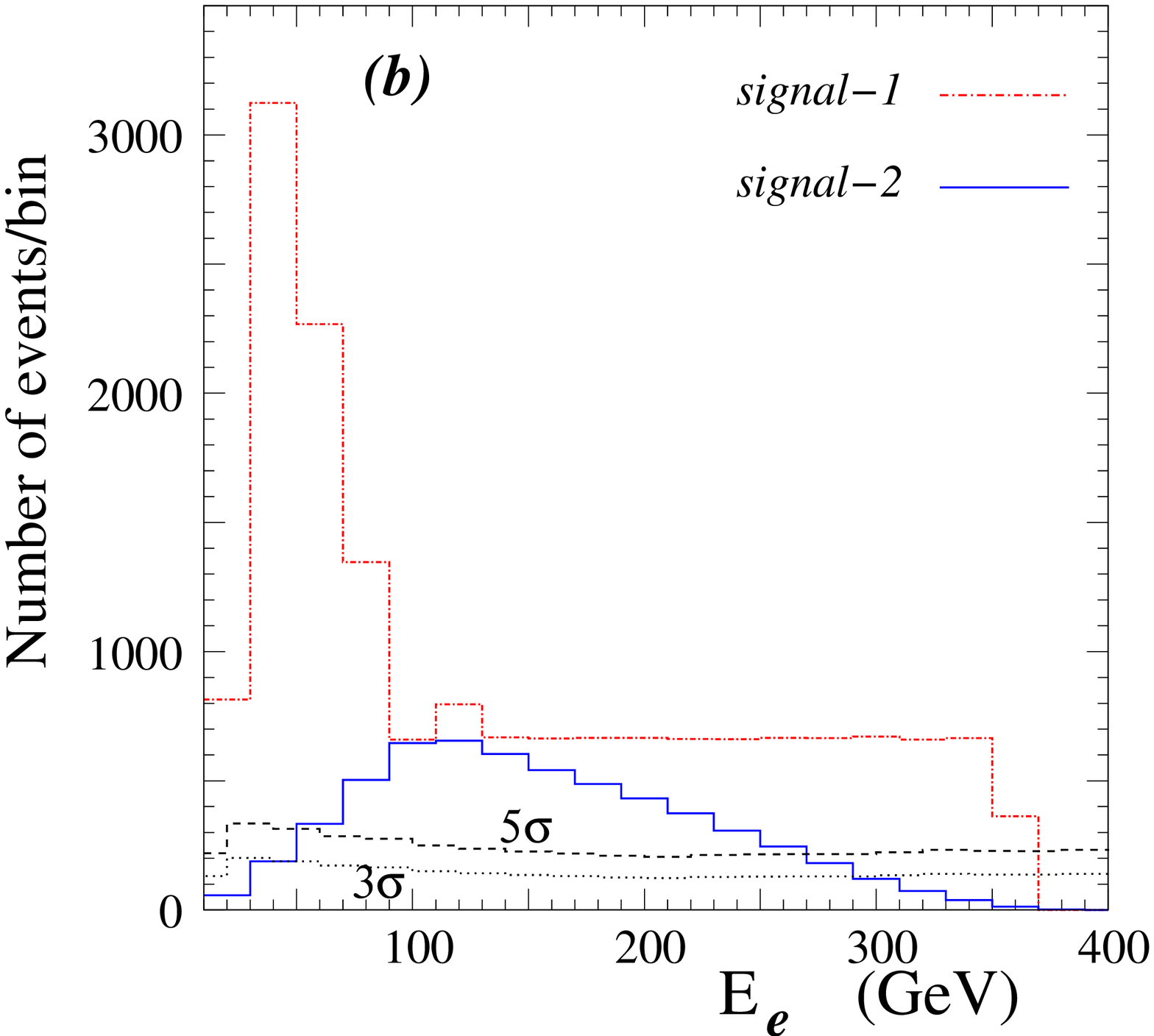}
\includegraphics[height=2.3in,width=2.3in]{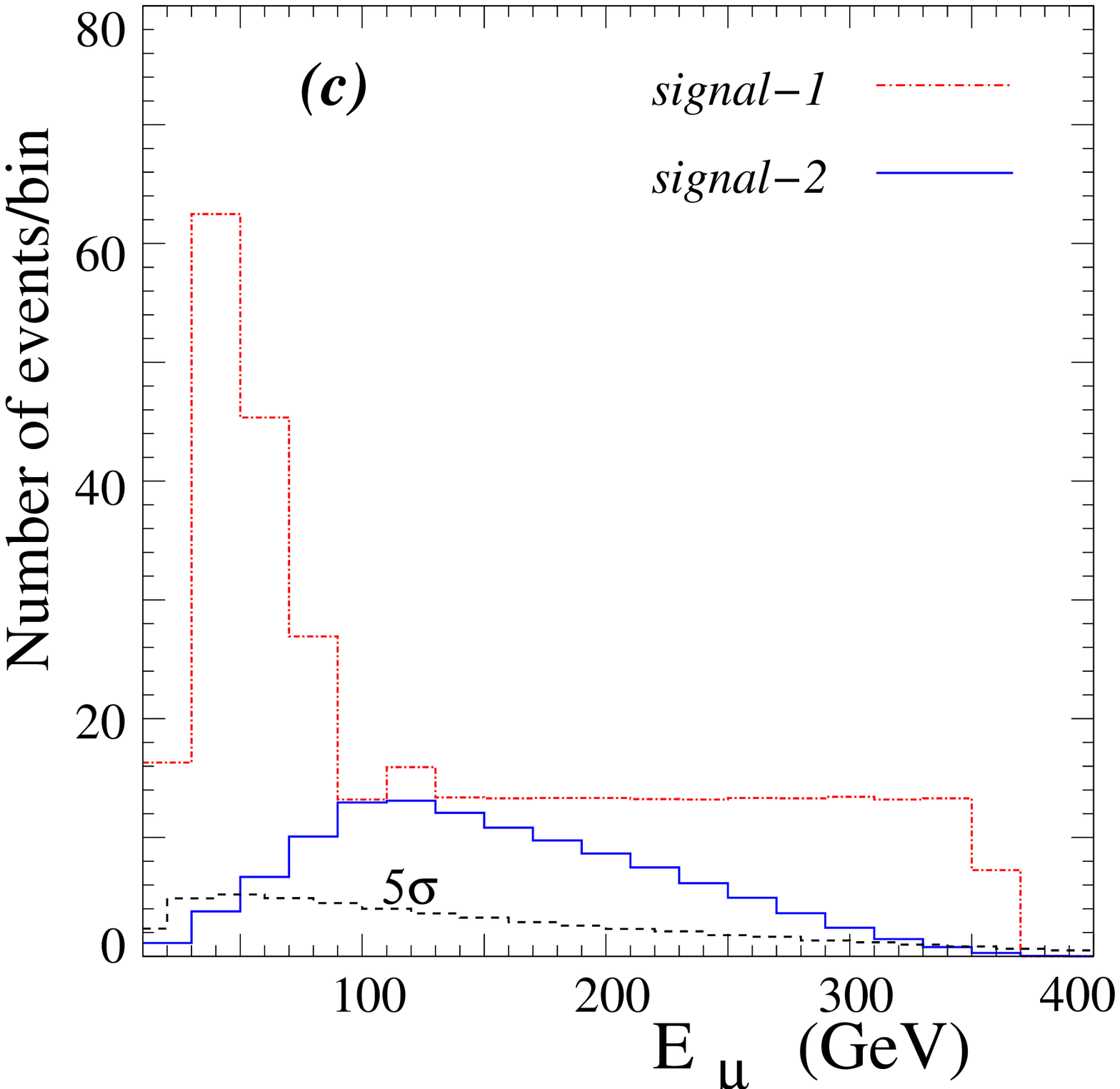}
\caption{\sl\small Binwise distribution of the energy of $e^-(\mu^-)$. 
The background follows the notation of Fig~\ref{energymiss1}. Each binsize is 
20 GeV ($\sqrt{s}=1$ TeV). 
(a) ${\cal L}=100 fb^{-1}$ (final state $e^-e^-E\slash$),
(b) ${\cal L}=500 fb^{-1}$  (final state $e^-e^-E\slash$), 
(c) ${\cal L}=10 fb^{-1}$  (final state $\mu^-\mu^-E\slash$). 
Here $\sqrt{s}=1$ TeV and $M_{\tilde\Delta_L^{--}}=$500 GeV.
For {\it signal-1} $m_{\tilde e_L}=$250 GeV  and
for {\it signal-2} $m_{\tilde e_L}=$500 GeV.}
\label{energydist2}
\end{center}
\end{figure}
{\it signal-1} the leptons coming from the 2-body decay of slepton will have 
an energy profile depending on the mass difference 
$m_{\tilde{\ell}}-M_{\N0_1}$,
while the lepton which comes from the 2-body decay of $\Dm$ will have an energy
profile dominantly depending on $M_{\Dm}-m_{\tilde{\ell}}$. 
For {\it signal-2} however, both leptons have identical distributions which
depend on the 3-body decay kinematics of the parent particle.

In Figs.~\ref{energymiss2}, \ref{invmass2} and \ref{energydist2} we show 
the distributions for the total missing energy, invariant mass
of the visible leptons and the energy profile of the final state leptons for 
the machine with $\sqrt{s}=1$ TeV center-of-mass energy for the {\it sample 
point} {\bf B} given in Table~\ref{susyin}. The mass for $\Dm_L$ is taken as 
500 GeV while the corresponding masses of the selectrons 
for {\it signal-1} and {\it signal-2} are 250 GeV and 550 GeV respectively. 
The contribution to {\it signal-1} from $\tilde{e}^-_L$-pair production in 
this case is however very small ($0.2~fb$) and does have any significant effect
on the distributions.
In each of the plots we also show the 
distributions for the final states $\mu^-\mu^-E\slash$. Here, in addition 
to distributions given for integrated luminosities ${\cal L}=100 fb^{-1}$ 
and ${\cal L}=500 fb^{-1}$, we also include the case (c) ${\cal L}=10 fb^{-1}$. 
The signal is quite large for heavier $\Dm_L$ and the signal far overwhelms the 
background for the $\mu^-\mu^-E\slash$. 
It can be seen in all the Figs.~\ref{energymiss2},\ref{invmass2} and 
\ref{energydist2} that we still have a large SM background and that an integrated 
luminosity of ${\cal L}=500 fb^{-1}$ is needed for a significant 
$5\sigma$-signal for
{\it signal-2} for the  $e^-e^-E\slash$ final state. The SM background for the
$\mu^-\mu^-E\slash$ final state is however still quite suppressed and one 
can observe the large rates for both {\it signal-1} and {\it signal-2} as compared 
to the $5\sigma$ statistical fluctuations in the SM for ${\cal L}=10  fb^{-1}$. 
We preferred to show the 
distributions on a linear scale as opposed to the earlier plots, since the 
rates for both {\it signal-1} and {\it signal-2} do not differ by a large 
factor. 
The distributions do not 
exhibit any additional new features as compared to the plots for the same 
kinematic variables at the $\sqrt{s}=500$ GeV machine. However a look at 
Table~\ref{rates} suggests that a heavier $\Dm$ accessible at a 
$\sqrt{s}=1$ TeV machine will have a substantial SM background and the
signal-background analysis merits discussion. 
\subsection{The right-handed higgsino $\Dm_R$}
We briefly discuss the production and decay of the right chiral higgsino 
\begin{figure}[htb]
\begin{center}
\includegraphics[height=3.2in,width=3.2in]{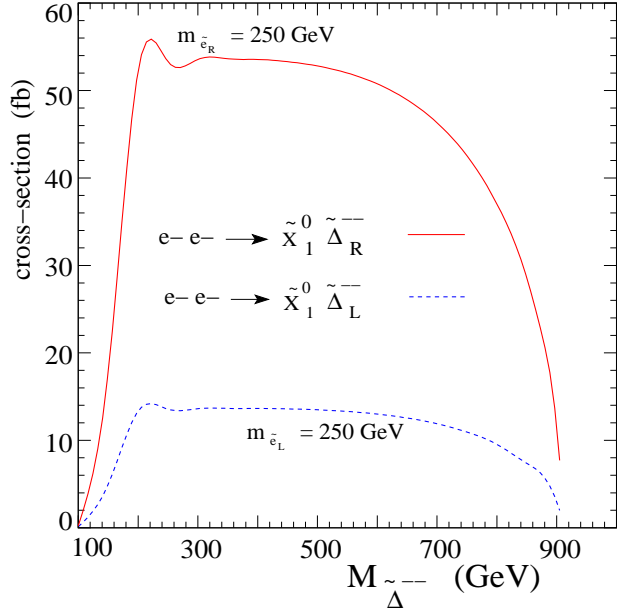}
\caption{\sl\small 
The production cross section as a function of the triplet higgsino
mass ($\sqrt{s}=1$ TeV). 
The cross section for both $\Dm_L$ and $\Dm_R$. The solid (red) line
corresponds to $\Dm_R$ production with the beam polarizations  (+1,+1) while
the broken (blue) line corresponds to $\Dm_L$ production with the beam 
polarizations (-1,-1). The plots shown are for the {\it sample point} {\bf A} 
given in Table~\ref{susyin}.}
\label{prodR}
\end{center}
\end{figure}
$\Dm_R$ at the $e^-e^-$ collider, emphasizing distinguishing features with respect 
to $\Dm_L$. In Fig.~\ref{prodR} we show the production cross section for 
its production and compare it with the production cross section of $\Dm_L$.
Note that the production of the right-chiral $\Dm_R$ requires the electron beams to 
be dominantly right-polarized and can be quite large compared to the production
cross section of the $\Dm_L$, as shown in Fig.~\ref{prodR}.
The difference is mainly due to the difference in coupling of the neutralino 
with $e-\widetilde{e}_{L(R)}$ which in this case reads, 
\beas
&&\ell^-{\tilde \ell}_R^{-} \chi_k^0~~\to~~\frac{1}{\sqrt{2}}(g_R N^*_{k2} 
+ g_V N^*_{k3}) P_R
\eeas
The production of the right-chiral
$\Dm_R$ will have a very clean signal as the SM background 
is completely reducible using polarization of the beams. We have already 
listed the SM background in Table~\ref{rates} corresponding to different 
beam polarizations. The right-polarized beams effectively kill the 
SM background because  the dominant contributions come from W-boson exchange
which vanish when the electron beams are right-polarized.  
So the signal for the $\Dm_R$ production is relatively background free.
In fact, this suggests that the signals for the right-handed $\Dm_R$ will
be even more striking and hence much lower values of the $\D L=2$ couplings can
be probed.
To highlight this fact and to illustrate the sensitivity of the signal
to the $\D L$ coupling, we calculate the rates for different values of the
$\D L$ coupling $\tilde{f}_{ee}$. We plot the distributions in the missing 
energy, invariant mass of the visible leptons and the energy of the visible 
lepton in the final state for the $e^-e^-E\slash$ final state, coming from the 
\begin{figure}[htb]
\begin{center}
\includegraphics[height=2.3in,width=2.3in]{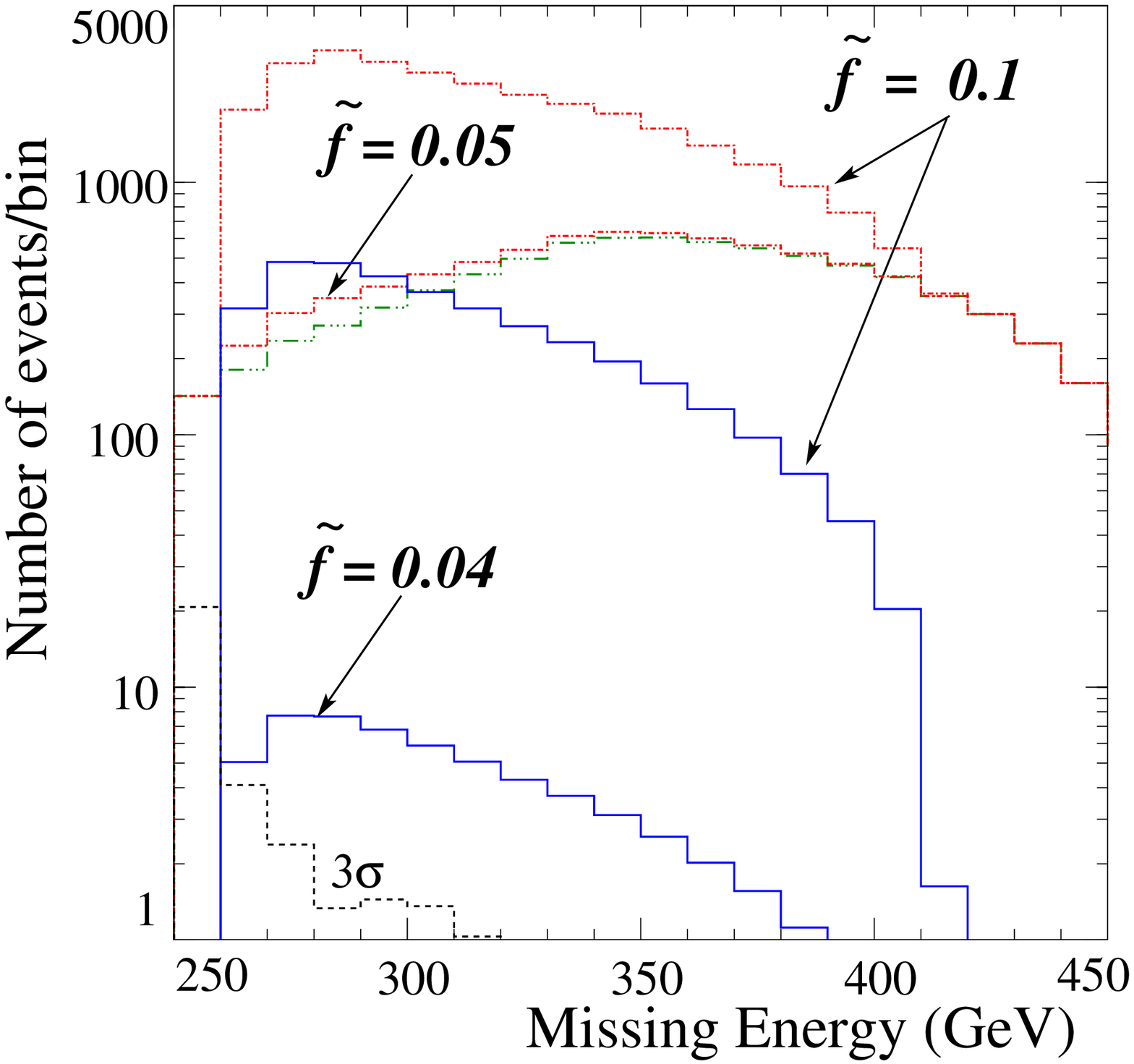}
\includegraphics[height=2.3in,width=2.3in]{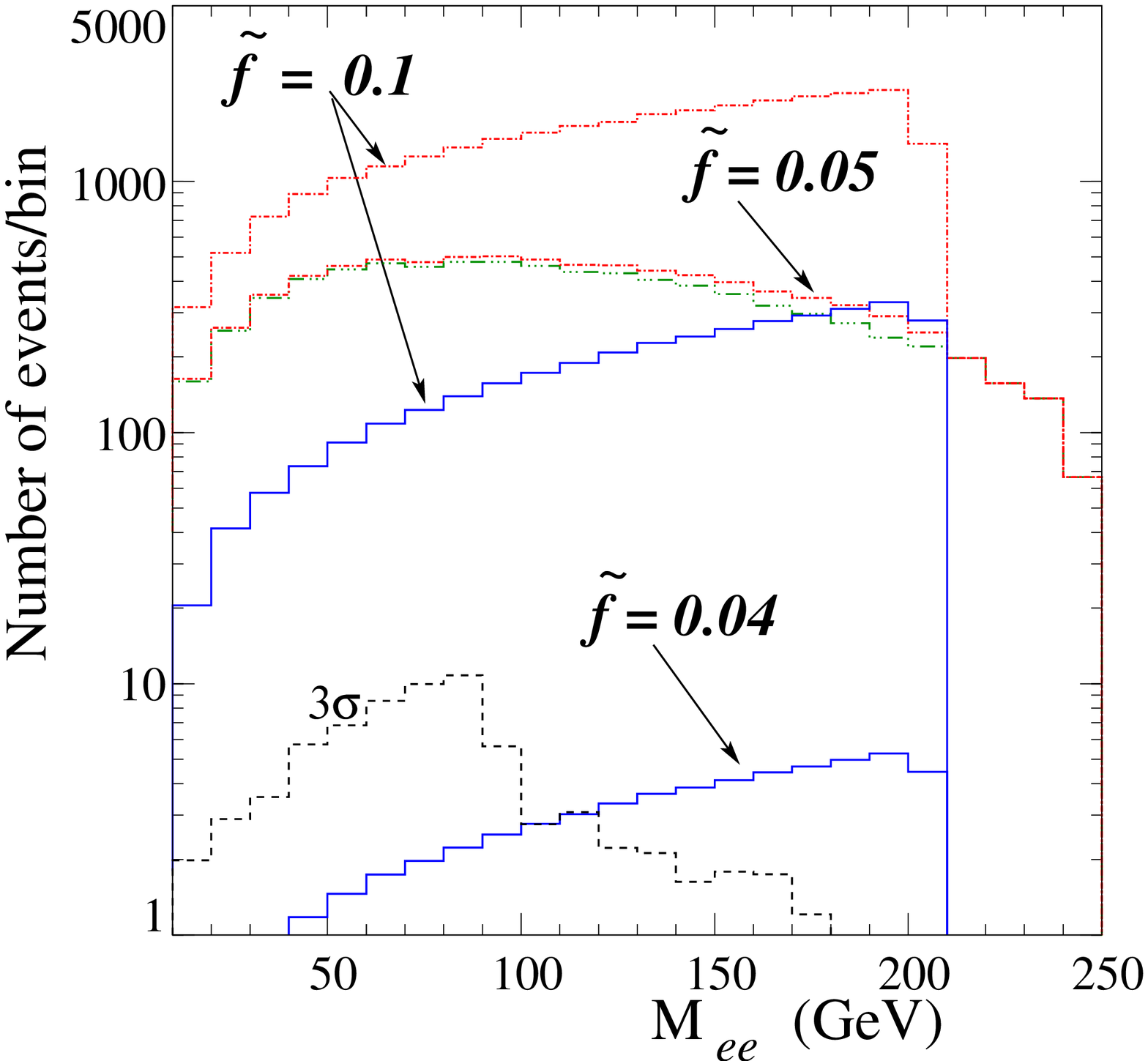}
\includegraphics[height=2.3in,width=2.3in]{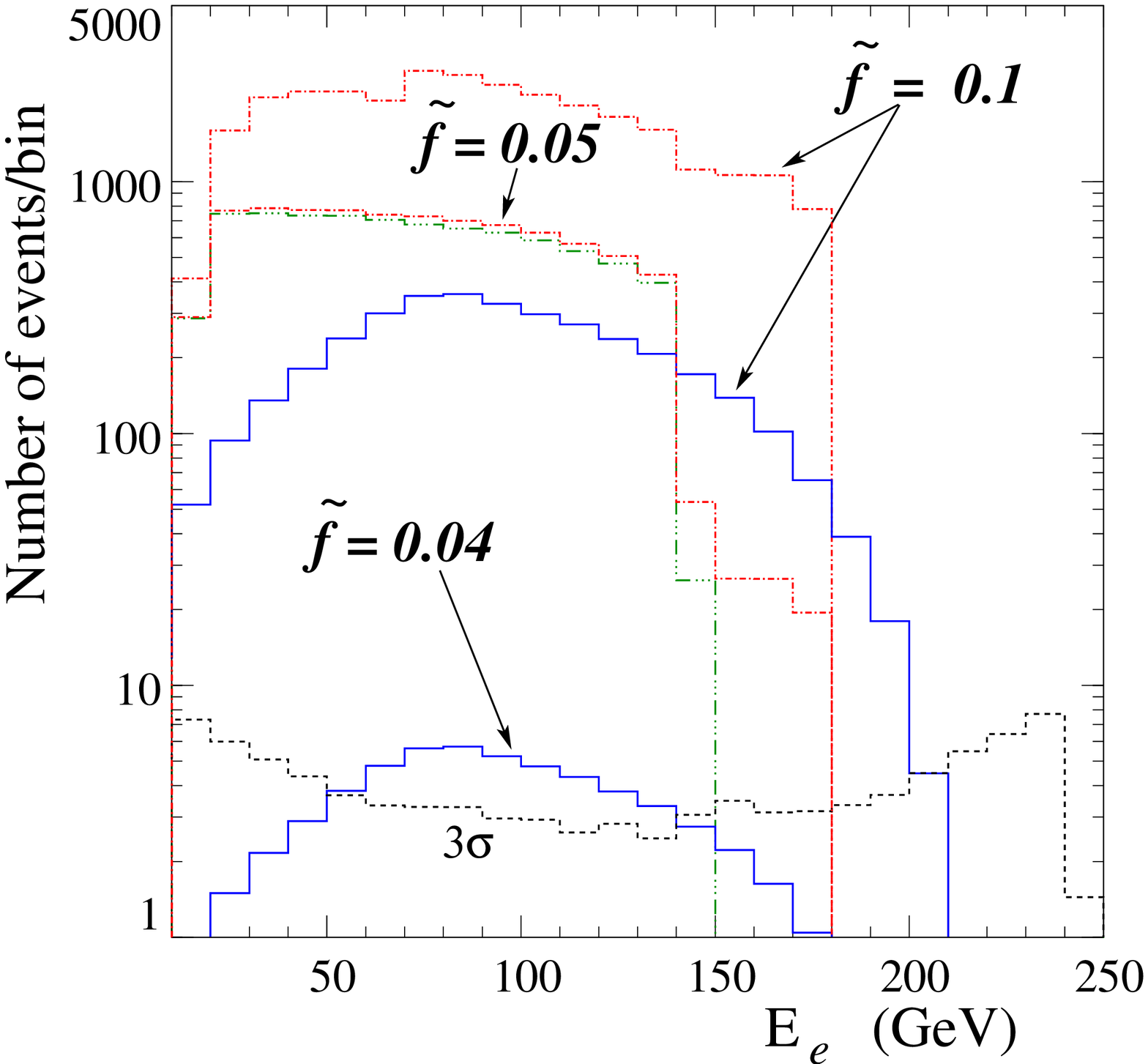}
\caption{\sl\small Binwise distribution of missing energy, invariant mass
and energy of $e^-$ for different values of the $\D L=2$ coupling 
$\tilde{f}_{ee}$. 
The statistical fluctuations in the SM background is shown at $3\sigma$
in black (dashed) lines. Each binsize is 10 GeV  while the luminosity is 
taken as ${\cal L}=500 fb^{-1}$ (final state $e^-e^-E\slash$).
The green ($-\cdot\cdot-$) line stands for the independent contribution
from $\tilde{e}^-_R$-pair production for {\it signal-1}.
Here $\sqrt{s}=500$ GeV and $M_{\tilde\Delta_R^{--}}=$300 GeV.
For {\it signal-1} shown in red lines, $m_{\tilde e_R}=$150 GeV  and
for {\it signal-2 shown in blue lines, $m_{\tilde e_R}=$400 GeV.}}
\label{distbn:right-higgsino}
\end{center}
\end{figure}
$\tilde{e}^-_R$-pair production and $\Dm_R$ production, and the subsequent 
decays in Fig~\ref{distbn:right-higgsino},
for three values of $\tilde{f}_{ee}=$0.04, 0.05 and 0.1. We consider a 
luminosity of ${\cal L}=500 fb^{-1}$ and show the $3\sigma$ fluctuations in the 
SM background. All the figures clearly show that {\it signal-1} stands out 
against the background for a $\D L$ coupling strength which can be as low as 
$\tilde{f}_{ee}=0.05$ and even lower values can be accessed if we have 
$\mu^-\mu^-E\slash$ final states. The green ($-\cdot\cdot-$) line stands for 
the independent contribution coming from $\tilde{e}^-_R$-pair production for 
{\it signal-1}. For the polarization choice (+1,+1), the cross section becomes 
$15.8~fb$. One can see that for $\tilde{f}_{ee} \to 0$ the distribution for 
{\it signal-1} will converge on the green ($-\cdot\cdot-$) line as the 
contribution coming from the $\Dm_R$ production becomes negligibly small.
The plot shown for {\it signal-2}, however shows the direct reach for the 
coupling $\tilde{f}_{ee}$ as the selectron-pair production is kinematically 
disfavored for a right-selectron mass of 400 GeV.  
 
\section{Summary and Conclusions}\label{sec:summ}
The absence of $\D L=2$ processes in the SM provides an unique platform 
to search for beyond the SM scenarios. Such processes arise naturally at an 
$e^- e^-$ collider where they can be produced directly.
LRSUSY models where R-parity is conserved, give rise to such doubly 
charged exotics which can be relatively light and accessible at future linear 
accelerators. In this work, we have studied in detail the production 
of a single doubly charged higgsino and its decay channels at the ILC 
operating in the $e^-e^-$ mode. 
We showed that the production cross sections for the $\Dm$ can be quite large 
if the beams are polarized and that enough statistics is available to study 
the signals arising from its subsequent decays against the dominant SM 
background.
We have also shown how kinematic cuts can effectively limit this background and 
give a clear signal for LRSUSY at an $e^-e^-$ collider. We described how 
different final states ($e^-e^-E\slash,~\mu^-\mu^-E\slash$) compare against
the dominant SM background. The use of polarized beams helps in producing
doubly charged higgsinos of different chirality and hence proves to be an 
essential tool in distinguishing between the two chiral states of the $\Dm$ 
producing the same final states. We also find that the production of the right 
chiral higgsino $\Dm_R$ can be relatively background free and is more sensitive 
to lower values of the  $\D L=2$ coupling. This is a major advantage over the 
production of these exotics at hadron colliders or in  $e^+e^-$ collisions 
where ascertaining the chiral nature of $\Dm$ will be a non-trivial issue. 
A dominantly right (left)-polarized option for both the electron beams can 
achieve this quite easily at the $e^-e^-$ collider. As most studies of LRSUSY 
expect the doubly-charged higgsinos to be light, and even allow for the next 
to lightest supersymmetric particle (NLSP) to be the right-handed doubly 
charged higgsino \cite{Chacko:1997cm}, the resulting signals could be 
spectacular.

To summarize, in this work we have shown that an $e^-e^-$ collider can be 
an ideal machine to study the production of doubly charged higgsinos of the 
LRSUSY model. Through the choice of a few representative points in the 
parameter space of the model, we have shown signals coming from the production 
of $\Dm$ for two center-of-mass energies $\sqrt{s}=500$ GeV and 
$\sqrt{s}=1$ TeV. We have shown that through the final states 
$\ell^-\ell^-E\slash$ we have clear signals for these exotics. We also find 
that when $\ell=\mu,\tau$ the SM background is a six-body final state and 
hugely suppressed. We have also highlighted the advantage of using polarized 
beams in producing these exotics and suppressing the SM background. 

\vskip 5pt
{\bf Acknowledgments}

The authors would like to thank F. Maltoni for useful technical guidance 
regarding the use of Madgraph. The work of MF was funded by NSERC of Canada 
under the grant number 0105354. KH and SKR greatfully acknowledge support from 
the Academy of Finland (project number 115032).


\section{Appendix}

In what follows, we use the conventions of \cite{Frank:2003un}.
\subsection{Charginos}

In the supersymmetric sector of the model there are four doubly charged
higgsinos $\tilde \Delta_L^{--}, \tilde \delta_L^{++},\tilde
\Delta_R^{--}$, and $\tilde \delta_R^{++}$. The mass terms relevant to
these higgsinos are:
\begin{equation}
{\cal L}_{\tilde \Delta}=-M_{\tilde \Delta^{--}}\tilde \Delta_L^{--}\tilde \delta_L^{++}-M_{\tilde \Delta^{--}}\tilde \Delta_R^{--}\tilde\delta_R^{++} +h.c., 
\end{equation}
where in our notation $M_{\tilde \Delta^{--}}=\mu_3$. The model also has six singly-charged charginos, corresponding to
$\tilde\lambda_{L}$, $\tilde\lambda_{R}$, $\tilde\phi_{u}$,
$\tilde\phi_{d}$, $\tilde\Delta_{L}^{\pm}$, and $\tilde\Delta_{R}^{\pm}$.
 
The terms relevant to the masses of charginos in the Lagrangian are
\begin{equation}
{\cal L}_C=-\frac{1}{2}(\psi^{+T}, \psi^{-T}) \left ( \begin{array}{cc}
                                                        0 & X^T \\
                                                        X & 0
                                                      \end{array}
                                              \right ) \left (
\begin{array}{c}
                                                               \psi^+ \\
                                                               \psi^-
                                                               \end{array}
                                                        \right ) + {\rm H.c.} \ ,
\end{equation}
where $\psi^{+T}=(-i \lambda^+_L, -i \lambda^+_R, \tilde{\phi}_{1d}^+,
\tilde{\phi}_{1u}^+, \tilde{\Delta}_L^+, \tilde{\Delta}_R^+)$
and $\psi^{-T}=(-i \lambda^-_L, -i \lambda^-_R, \tilde{\phi}_{2d}^-,
\tilde{\phi}_{2u}^-, \tilde{\delta}_L^-, \tilde{\delta}_R^-)$, and
\begin{equation}
X=\left( \begin{array}{cccccc}
                            M_L & 0 & 0 & g_L\kappa_d & \sqrt{2}g_Lv_{\delta_L} &0
\\
           0 & M_R &0  & g_R\kappa_d &0
&\sqrt{2}g_Rv_{\delta_R}
\\
    g_L\kappa_u & g_R \kappa_u & 0 &-\mu_1 &  0 &0
\\
0 & 0  & -\mu_1 &0 & 0&0\\
 \sqrt{2} g_L v_{\Delta_L} & 0 & 0 &0 & -\mu_3 &0 \\
       0 & \sqrt{2} g_R v_{\Delta_R} & 0 & 0 &0& -\mu_3
               \end{array}
         \right ),
\end{equation}
where we have taken, for simplification, $\mu_{ij}=\mu_1$. Here $\kappa_u$
and $\kappa_d$ are the bidoublet Higgs bosons vacuum expectation values
(VEVs), $v_{\Delta_R}$ and $v_{\delta_R}$ are the triplet Higgs bosons
VEVs, and $M_{L}, M_{R}$ the $SU(2)_L$ and $SU(2)_R$ gaugino masses,
respectively. The chargino mass eigenstates $\chi_i$ are obtained by
\begin{eqnarray}
\chi_i^+=V_{ij}\psi_j^+, \ \chi_i^-=U_{ij}\psi_j^-, \ i,j=1, \ldots 5,
\end{eqnarray}
with $V$ and $U$ unitary matrices satisfying
\begin{equation}
U^* X V^{-1} = M_D.
\end{equation}
The diagonalizing matrices $U^*$ and $V$ are obtained by
computing the eigenvectors corresponding
to the eigenvalues of $X X^{\dagger}$ and $X^{\dagger} X$, respectively.

\subsection{Neutralinos}
\label{subsec:neutralinos}

The model has eleven neutralinos, corresponding to
$\tilde\lambda_{Z}$,
$\tilde\lambda_{Z^{\prime}}$,
$\tilde\lambda_{B-L}$,  $\tilde\phi_{1u}^0$, $\tilde\phi_{2u}^0$,
$\tilde\phi_{1d}^0$,  $\tilde\phi_{2d}^0$, $\tilde\Delta_{L}^0$,
$\tilde\Delta_{R}^0$,  $\tilde\delta_{L}^0$, and
$\tilde\delta_{R}^0$. 
The terms relevant to the masses of neutralinos in the Lagrangian are
\begin{equation}
{\cal L}_N=-\frac{1}{2} {\psi^0}^T Y \psi^0  + H.c. \ ,
\end{equation}
where $\psi^0=(-i \lambda_L^0, -i \lambda_R^0, -i \lambda_{B-L},
\tilde{\phi}_{1u}^0, \tilde{\phi}^0_{2d},\tilde{\Delta}_L^0,
\tilde{\delta}_L^0, \tilde{\Delta}_R^0, \tilde{\delta}_R^0)^T$
We omit $(\tilde{\phi}_{1d}^0, \tilde{\phi}^0_{2u} )^T $, which only mix
with each other.
The mixing matrix is:
\begin{equation}
\displaystyle
Z\!\!=\!\!\!\left(\!\! \begin{array}{ccccccccc}
              \!M_{L} &\! 0 & \!0 &\! -\frac{g_L \kappa_u}{\sqrt{2}} &\!
\frac{g_L \kappa_d}{\sqrt{2}} &\! -2^{\frac12}g_Lv_{\Delta_L}&\! -2^{\frac12}g_Lv_{\delta_L}&\!0 &\! 0 
\\
         \!0 & \!M_{R} &\! 0 &\! \frac{g_L \kappa_u}{\sqrt{2}} &\! \frac{g_L
\kappa_d}{\sqrt{2}} &\! 0 &\!0 &\! -2^{\frac12}g_Rv_{\Delta_R} &\! -2^{\frac12}g_R v_{\delta_R}
 \\
           \!0 \!& \!0 & \!M_{B-L} & \!0 & \!0 & \!2^{\frac32}g_V v_{\Delta_L} & \!2^{\frac32} g_V
v_{\delta_L} &\!2^{\frac32}g_V v_{\Delta_R} &\! 2^{\frac32} g_V
v_{\delta_R}
\\
        \! -\frac{g_L \kappa_u}{\sqrt{2}}&\! \frac{g_R \kappa_u}{\sqrt{2}}
&\! 0 &\! 0 & \!\mu_1 &\! 0 &\! 0 &\! 0 &\! 0  
\\
             \! \frac{g_L \kappa_d}{\sqrt{2}} &\! -\frac{g_R \kappa_d}{\sqrt{2}} & \!0 &\!\mu_1
& \!0 & \!0 &\! 0 & \!0 & \!0  \\
\!-2^{\frac12}g_Lv_{\Delta_L} & \!0 &\!2^{\frac32}g_Vv_{\Delta_L}&\! 0&\! 0&\!0 &\!-\mu_3 &\! 0 &\!0 \\
  \!-2^{\frac12}g_L v_{\delta_L} &\! 0& \! 2^{\frac32}g_V v_{\delta_L} & \!0  &\!0 &\! -\mu_3 & \!0 &\!0 &\!0
  \\
             \!0&\!-2^{\frac12}g_Rv_{\Delta_R} &\!2^{\frac32}g_Vv_{\Delta_R}& \!0&\! 0&\! 0 & \!0 &\!0 &\!-\mu_3 
\\
\! 0 &\! -2^{\frac12}g_R v_{\delta_R} &\! 2^{\frac32}g_V v_{\delta_R} & \!0 & \!0&\! 0 &\! 0 &\! -\mu_3 &\! 0 
       \end{array}\!\! \!
        \right )
\end{equation}
The mass eigenstates are defined by
\begin{equation}
\chi^0_i=N_{ij} \psi^0_j \ (i,j=1,2, \ldots 9),
\end{equation}
where $N$ is a unitary matrix chosen such that
\begin{equation}
N ZN^T = Z_D,
\label{equationN}
\end{equation}
and $Z_D$ is a diagonal matrix with non-negative entries. 

\subsection{Scalars}

The interactions between vector bosons and scalars arise from the kinetic
energy term for the gauge bosons in the Lagrangian density. We denote by
$x_{L,R}$ the Higgs fields before mixing, and by $y_{L,R}$ the Higgs
fields after mixing. The Higgs scalar fields are defined as:

\noindent
{\it Doubly Charged Fields}
\begin{eqnarray}
x_R^{++T} & \equiv & \left ( \Delta_R^{++} ~~\delta_R^{-- \ast}\right ),~
x_R^{--T}  \equiv \left ( \Delta_R^{++\ast} ~~\delta_R^{--}\right )
\nonumber \\ 
y_R^{\pm \pm T} & \equiv & \left (H_1^{\pm \pm} ~~H_2^{\pm \pm} \right )
\nonumber \\
x_L^{++T} & \equiv & \left ( \Delta_L^{++} ~~\delta_L^{-- \ast}\right ),~
x_L^{--T}  \equiv \left ( \Delta_L^{++\ast} ~~\delta_L^{--}\right )
\nonumber \\ 
y_L^{\pm \pm T} & \equiv & \left (H_3^{\pm \pm} ~~H_4^{\pm \pm} \right )
\end{eqnarray}
{\it Singly Charged Fields}
\begin{eqnarray}
x^{+T} &\equiv & \left ( \Delta_L^+~~\delta_L^{-
\ast}~~\phi_{2d}^{-\ast}~~\phi_{1u}^+~~\phi_{2u}^{-\ast}~~\phi_{1d}^+~~\Delta_R^+~~\delta_R^{-\ast}
\right), \nonumber \\
x^{-T} &\equiv & \left (
\Delta_L^{+\ast}~~\delta_L^{-}~~\phi_{2d}^{-}~~\phi_{1u}^{+\ast}~~
\phi_{2u}^{-}~~\phi_{1d}^{+ \ast}~~\Delta_R^{+\ast}~~\delta_R^{-}
\right), \nonumber \\
y^{\pm T} &\equiv & \left
(H_1^\pm~~H_2^\pm~~H_3^\pm~~H_4^\pm~~H_5^\pm~~H_6^\pm~~
G_1^\pm~~G_2^\pm\right)
\end{eqnarray}
{\it Neutral Fields}
\begin{eqnarray}
x_s^{0T} &\equiv & \left (
H_{\Delta_L}~~H_{\delta_L}~~H_{1d}^{0}~~H_{2u}^0~~H_{1u}^{0}~~H_{2d}^0~~
H_{\Delta_R}~~H_{\delta_R}
\right), \nonumber \\
y_s^{0 T} &\equiv & \left
(H_1^0~~H_2^0~~H_3^0~~H_4^0~~H_5^0~~H_6^0~~
H_7^0~~H_8^0 \right)
, \nonumber \\
x_p^{0T} &\equiv & \left (
z_{\Delta_L}~~z_{\delta_L}~~z_{1d}^{0}~~z_{2u}^0~~z_{1u}^{0}~~z_{2d}^0~~
z_{\Delta_R}~~z_{\delta_R}
\right), \nonumber \\
y_p^{0 T} &\equiv & \left
(A_1^0~~A_2^0~~A_3^0~~A_4^0~~A_5^0~~A_6^0~~
G_1^0~~G_2^0 \right)
\end{eqnarray}
where the indeces "$s$" and "$p$" stand for scalar and pseudoscalar,
respectively. There are two charged Goldstone bosons for the left-handed
and the right-handed charged vector bosons, and two neutral Goldstone
bosons for the $Z_L$ and $Z_R$ bosons. They have zero mass. We define
them to be the 7-th and 8-th component of $H^\pm$ and $A^0$, in order to
simplify the summation convention. The mass matrices $M$
are real and symmetric and diagonalized by orthogonal matrices
$R$ defines as:
\begin{eqnarray}
\left (R_R^{\pm \pm} \right )_{ij} \left ( M_R^{\pm \pm} \right
)_{jk}\left (R_R^{\pm \pm} \right )_{lk}&=& \mbox {diag} \left (m_1^{\pm
\pm}, m_2^{\pm \pm} \right ),\nonumber \\
\left (R_L^{\pm \pm} \right )_{ij} \left ( M_L^{\pm \pm} \right
)_{jk}\left (R_L^{\pm \pm} \right )_{lk}&=& \mbox {diag} \left (m_3^{\pm
\pm}, m_4^{\pm \pm} \right ),\nonumber \\
\left (R^{\pm } \right )_{ij} \left ( M^{\pm } \right
)_{jk}\left (R^{\pm} \right )_{lk}&=& \mbox {diag} \left
(m_1^{\pm},\ldots , m_6^{\pm},0,0 \right ),\nonumber \\
\left (R_s^{0} \right )_{ij} \left ( M_s^{0} \right
)_{jk}\left (R_s^{0} \right )_{lk}&=& \mbox {diag} \left (m_{s1}^{0},
\ldots ,m_{s8}^{0} \right ),\nonumber \\
\left (R_p^{0} \right )_{ij} \left ( M_p^{0} \right
)_{jk}\left (R_p^{0} \right )_{lk}&=& \mbox {diag} \left (m_{p1}^{0},
\ldots ,m_{s6}^{0},0,0 \right )
\end{eqnarray}
where $m_{s1}$ is the mass of the lightest Higgs scalar. 
Diagonalizing the scalar mass matrices, we
introduce new fields by:
\begin{eqnarray}
y_{Ri}^{\pm \pm}&=&\left (R_R^{\pm \pm}\right )_{ij}x_{Rj}^{\pm \pm},~~
y_{Li}^{\pm \pm}=\left (R_L^{\pm \pm}\right )_{ij}x_{Lj}^{\pm
\pm},~~y_i^\pm=\left ( R^\pm \right )_{ij}x^\pm_j,
\nonumber \\
y_{si}^{0}&=&\left (R_s^{0}\right )_{ij}x_{s j}^{0},~~
y_{pi}^{0}=\left (R_p^{0}\right )_{ij}x_{pj}^{0}
\end{eqnarray}

\subsection{Couplings}
\label{subsec:couplings}

Finally, we list the all couplings involving doubly charged higgsinos
relevant for the production and decay:
\begin{eqnarray}
&&{\tilde \Delta_R}^{--}{\Delta_R}^{--} \chi_k^0~~~~~\left (\frac{e}{\sqrt{\cos 2 \theta_W}} N_{k1}+gN_{k3} \right )P_R
\\
&&{\tilde \Delta_L}^{--}{\Delta_L}^{--} \chi_k^0~~~~~\left (\frac{e}{\sqrt{\cos 2 \theta_W}}N_{k1}+gN_{k2}\right ) P_L
\\
&&{\tilde \Delta_R}^{--}{\Delta_R}_j^{-} \chi_k^- ~~~~~~~~gU_{k2}A_{j7}P_R
\\
&&{\tilde \Delta_L}^{--}{\Delta_L}_j^{-} \chi_k^-~~~~~~~~gU_{k2}A_{j1}P_L
\\
&&{\tilde \Delta_R}^{--}{\tilde l_R}^{-} l_R^-~~~~~~~~~~-2h_{ll}\mathcal{C}^{-1}P_R \\
&&{\tilde \Delta_L}^{--}{\tilde l_L}^{-} l_L^-~~~~~~~~~~-2h_{ll}\mathcal{C}^{-1}
P_L
\end{eqnarray}
where $A_{ij}$ is the matrix which diagonalizes the mass matrix for the
singly-charged Higgs bosons. $\mathcal{C}$ is the charge conjugation
operator while $P_L$ and $P_R$ are the chirality projection operators
$(1\mp \gamma_5)/2$.



\begin{thebibliography}{99}

\bibitem{Eidelman:2004wy}
  S.~Eidelman {\it et al.}  [Particle Data Group],
  Phys.\ Lett.\ B {\bf 592} (2004) 1.

\bibitem{Spergel:2003cb}
  D.~N.~Spergel {\it et al.}  [WMAP Collaboration],
  Astrophys.\ J.\ Suppl.\  {\bf 148}, 175 (2003).

\bibitem{Barbier:2004ez}
  R.~Barbier {\it et al.},
  Phys.\ Rept.\  {\bf 420} (2005) 1.
  
    
\bibitem{Mohapatra:1979ia}
  R.~N.~Mohapatra and G.~Senjanovic,
  Phys.\ Rev.\ Lett.\  {\bf 44}, 912 (1980);
  R.~N.~Mohapatra and G.~Senjanovic,
  Phys.\ Rev.\ D {\bf 23}, 165 (1981); 
M. Gell-Mann, P. Ramond, R. Slansky, Supergravity (P. van Nieuwenhuizen
et al. eds.), North Holland, Amsterdam, 1980, p. 315; 
T. Yanagida, in Proceedings of the Workshop on the Unified
Theory and the Baryon Number in the Universe (O. Sawada and
A. Sugamoto, eds.), KEK, Tsukuba, Japan, 1979, p. 95. 
S.L. Glashow, The future of elementary particle physics, in
Proceedings of the Summer Institute on Quarks and Leptons (M. Levy
et al eds.), Plenum Press, New York, 1980, pp. 687.

 
\bibitem{history} M. Cveti\v{c} and J. Pati, Phys.\ Lett.\ B {\bf 135}, 57 (1984); R. N. Mohapatra and A. Ra\v{s}in, Phys.\ Rev.\ D {\bf 54}, 5835 (1996); R. Kuchimanchi,  Phys.\ Rev.\ Lett. {\bf 76}, 3486 (1996); R. N. Mohapatra, A. Ra\v{s}in and  G. Senjanovi{\'c},  Phys.\ Rev.\ Lett. {\bf 79}, 4744 (1997); C. S. Aulakh, K. Benakli and G. Senjanovi{\'c}, Phys.\ Rev.\ Lett. {\bf79}, 2188 (1997); C. S. Aulakh, A. Melfo and G. Senjanovi{\'c}, Phys.\ Rev.\ D {\bf 57}, 4174 (1998).
 
\bibitem{Demir:2006ef}
  D.~A.~Demir, M.~Frank and I.~Turan,
  Phys.\ Rev.\  D {\bf 73}, 115001 (2006).

%

\bibitem{Francis:1990pi}
  R.~M.~Francis, M.~Frank and C.~S.~Kalman,
  Phys.\ Rev.\  D {\bf 43}, 2369 (1991);
  K.~Huitu and J.~Maalampi,
  Phys.\ Lett.\  B {\bf 344}, 217 (1995);
  K.~Huitu, J.~Maalampi and M.~Raidal,
  Phys.\ Lett.\  B {\bf 328}, 60 (1994);
  K.~Huitu, J.~Maalampi and M.~Raidal,
  Nucl.\ Phys.\  B {\bf 420}, 449 (1994).

 \bibitem{Mohapatra:1995xd}  
 R.~N.~Mohapatra and A.~Rasin,  
 Phys.\ Rev.\ Lett.\  {\bf 76}, 3490 (1996); 
  R.~N.~Mohapatra and A.~Rasin, 
   Phys.\ Rev.\ D {\bf 54}, 5835 (1996);  
      R.~Kuchimanchi,  
      Phys.\ Rev.\ Lett.\  {\bf 76}, 3486 (1996).
      
  \bibitem{Chacko:1997cm}  
Z.~Chacko and R.~N.~Mohapatra,  
Phys.\ Rev.\ D {\bf 58}, 015003 (1998); 
B.~Dutta and R.~N.~Mohapatra,  
Phys.\ Rev.\ D {\bf 59}, 015018 (1999).
 
\bibitem{Barenboim:1996pt}
  G.~Barenboim, K.~Huitu, J.~Maalampi and M.~Raidal,
  Phys.\ Lett.\  B {\bf 394}, 132 (1997).
    
\bibitem{Huitu:1996su}
K.~Huitu, J.~Maalampi, A.~Pietila and M.~Raidal,
  Nucl.\ Phys.\ B {\bf 487}, 27 (1997).
  
\bibitem{Huitu:1995bc}
  K.~Huitu, J.~Maalampi and M.~Raidal, HU-SEFT-I-1995-1, 1995;
  M.~Frank,
  Phys.\ Rev.\ D {\bf 62}, 053004 (2000).
  
\bibitem{Raidal:1998vi}
  M.~Raidal and P.~M.~Zerwas,
  Eur.\ Phys.\ J.\  C {\bf 8}, 479 (1999).


\bibitem{Singer:1980sw}
  M.~Singer, J.~W.~F.~Valle and J.~Schechter,
  Phys.\ Rev.\  D {\bf 22}, 738 (1980);
  J.~C.~Montero, F.~Pisano and V.~Pleitez,
  Phys.\ Rev.\  D {\bf 47}, 2918 (1993);
  P.~H.~Frampton,
  Phys.\ Rev.\ Lett.\  {\bf 69}, 2889 (1992);
  R.~Foot, O.~F.~Hernandez, F.~Pisano and V.~Pleitez,
  Phys.\ Rev.\  D {\bf 47}, 4158 (1993):
  R.~Foot, H.~N.~Long and T.~A.~Tran,
  Phys.\ Rev.\  D {\bf 50}, 34 (1994);
  J.~C.~Montero, F.~Pisano and V.~Pleitez,
  Phys.\ Rev.\  D {\bf 47}, 2918 (1993);
  H.~N.~Long,
  Phys.\ Rev.\  D {\bf 54}, 4691 (1996).
\bibitem{Swartz:1989qz}
   M.~L.~Swartz,
   Phys.\ Rev.\  D {\bf 40} (1989) 1521;
   J.~F.~Gunion, J.~Grifols, A.~Mendez, B.~Kayser and F.~I.~Olness,
   Phys.\ Rev.\  D {\bf 40} (1989) 1546;
   M.~Lusignoli and S.~Petrarca,
   Phys.\ Lett.\  B {\bf 226} (1989) 397;


\bibitem{Willmann:1998gd}
  L.~Willmann {\it et al.},
  Phys.\ Rev.\ Lett.\  {\bf 82} (1999) 49
  [arXiv:hep-ex/9807011];
  D.~Chang and W.~Y.~Keung,
  Phys.\ Rev.\ Lett.\  {\bf 62} (1989) 2583:
   S.~Godfrey, P.~Kalyniak and N.~Romanenko,
   Phys.\ Rev.\  D {\bf 65} (2002) 033009
   [arXiv:hep-ph/0108258].


\bibitem{Abbiendi:2003pr}
   G.~Abbiendi {\it et al.}  [OPAL Collaboration],
   Phys.\ Lett.\  B {\bf 577} (2003) 93
   [arXiv:hep-ex/0308052].

\bibitem{Achard:2003mv}
  P.~Achard {\it et al.}  [L3 Collaboration],
  Phys.\ Lett.\  B {\bf 576} (2003) 18
  [arXiv:hep-ex/0309076].

\bibitem{Abdallah:2002qj}
  J.~Abdallah {\it et al.}  [DELPHI Collaboration],
  Phys.\ Lett.\  B {\bf 552} (2003) 127
  [arXiv:hep-ex/0303026].

\bibitem{Aktas:2006nu}
   A.~Aktas {\it et al.}  [H1 Collaboration],
   Phys.\ Lett.\  B {\bf 638} (2006) 432
   [arXiv:hep-ex/0604027].

\bibitem{Heuer:2006ka}
  R.~D.~Heuer,
  Nucl.\ Phys.\ Proc.\ Suppl.\  {\bf 154}, 131 (2006).

\bibitem{Maltoni:2002qb}
  F.~Maltoni and T.~Stelzer,
  JHEP {\bf 0302}, 027 (2003)
  [arXiv:hep-ph/0208156].

\bibitem{Frank:2003un}
  M.~Frank and P.~Pnevmonidis,
  Phys.\ Rev.\  D {\bf 67}, 015010 (2003).


\end{thebibliography}
\end{document}